\documentclass[11pt, letterpaper]{article}

\usepackage{graphicx}
\usepackage{amsmath}
\usepackage{amssymb}
\usepackage{setspace}
\usepackage{epstopdf}
\usepackage{color}
\usepackage[letterpaper,left=1in,right=1in,top=1in,bottom=1in]{geometry}
\usepackage{multirow}
\usepackage{longtable}
\usepackage{rotating}
\usepackage{empheq,etoolbox}
\usepackage{algpseudocode}
\usepackage{algorithm}

\usepackage{titlesec}

\allowdisplaybreaks

\title{\Large \bf Model predictive control of agro-hydrological systems based on a two-layer neural network modeling framework}
\author{
	\centerline{\normalsize Zhiyinan Huang$^{a}$, Jinfeng Liu$^{a,}$\thanks{Corresponding author: J. Liu. Tel: +1-780-492-1317. Email: jinfeng@ualberta.ca.}, Biao Huang$^{a}$}\\
	\vspace{5mm}\\
	\centerline{\small $^{a}$Department of Chemical \& Materials Engineering, University of Alberta,}\\
	\centerline{\small Edmonton, AB T6G 1H9, Canada}
}
\allowdisplaybreaks

\begin{document}
	\date{}
	\maketitle
	\setstretch{1.39}
	
	\begin{abstract}                
		Water scarcity is an urgent issue to be resolved and improving irrigation water-use efficiency through closed-loop control is essential. The complex agro-hydrological system dynamics, however, often pose challenges in closed-loop control applications. In this work, we propose a two-layer neural network (NN) framework to approximate the dynamics of the agro-hydrological system. To minimize the prediction error, a linear bias correction is added to the proposed model. The model is employed by a model predictive controller with zone tracking (ZMPC), which aims to keep the root zone soil moisture in the target zone while minimizing the total amount of irrigation.
		The performance of the proposed approximation model framework is shown to be better compared to a benchmark long-short-term-memory (LSTM) model for both open-loop and closed-loop applications. Significant computational cost reduction of the ZMPC is achieved with the proposed framework. To handle the tracking offset caused by the plant-model-mismatch of the proposed NN framework, a shrinking target zone is proposed for the ZMPC. Different hyper-parameters of the shrinking zone in the presence of noise and weather disturbances are investigated, of which the control performance is compared to a ZMPC with a time-invariant target zone.
	\end{abstract}

	
	\section{Introduction}
	Water scarcity is a rapidly escalating global issue due to various factors such as population growth and climate change. Approximately 70\% of the freshwater is consumed by agriculture activities \cite{guan_assessment_2007, mubako_inputoutput_2013}. Thus, improving the water-use efficiency in the agriculture industry is essential. Currently, practical irrigation policies are mostly open-loop, which are determined based on heuristic or empirical knowledge. Real-time feedback from the field is often not considered. This approach typically has low irrigation efficiency, as the applied irrigation amount can be imprecise, leading to over or insufficient irrigation.

	Increasing attentions have been drawn to closed-loop irrigation over the past decade \cite{kim_evaluation_2009, goodchild_method_2015}. Among various approaches, advanced control strategies are popular due to their ability of handling constraints and multiple objectives simultaneously \cite{saleem_model_2013, lozoya_model_2014, guo_data-driven_2018, mao_soil_2018, nahar_closed-loop_2019, sahoo_knowledge-based_2022}.  In \cite{saleem_model_2013, lozoya_model_2014, guo_data-driven_2018}, conventional model predictive control (MPC) is employed to optimize the soil-moisture dynamics in real-time in the presence of weather disturbances.
	Instead of tracking set-point, Mao et al. proposed an MPC controller that tracks a target zone (ZMPC), which provides more degrees of freedom in control actions \cite{mao_soil_2018}. Long-term irrigation scheduling that aims to optimize crop production over a longer time duration is investigated in \cite{nahar_closed-loop_2019, sahoo_knowledge-based_2022}. To be more specific, Nahar et al. proposed a hierarchical framework including a scheduler and a controller \cite{nahar_closed-loop_2019}. The scheduler has a larger sampling time and optimizes over the entire crop growth season. Some soil moisture reference is provided by the scheduler and is tracked by the controller over shorter horizons. Sahoo et al. proposed a knowledge-based scheduler that optimizes the irrigation amount and time explicitly and is shown to be effective under different scenarios \cite{sahoo_knowledge-based_2022}. 
	
	One of the major challenges in applying advanced control to the agro-hydrological system is the high computational cost. The water dynamics are often described by the Richards equation, which is a nonlinear PDE developed based on first principles \cite{richards_capillary_1931}. Spatial discretization is often required to simplify the Richards equation in applications \cite{richards_capillary_1931}, however, the resulting nonlinear ODE system is still difficult to solve and leads to states with higher dimensions. 
	
	Model reduction or approximation techniques are commonly used to reduce the computational complexity of high-dimensional nonlinear models. 
	Model reduction is carried out based on existing models and aims to simplify it. One of the widely used approaches is to project the high dimension original state space onto a lower dimension subspace \cite{ibrir_projection-based_2018, acle_parameter_2019}, of which a representative method is the proper orthogonal decomposition (POD) \cite{nguyen_pod-deim_2020}. However, the physical meanings of the original states are often not preserved by the projection approach, making it difficult to handle state constraints. 
	Data-based model identification method is often used to approximate complex nonlinear models \cite{mao_soil_2018, zhang_economic_2019, diao_event-triggered_2018, liu_rapid_2020} with no first-principle knowledge regarding the system required. The drawback, however, is that traditional model identification approaches require the model structure to be selected ahead of time, which may significantly affect the model performance. Furthermore, the computational cost of the identified model might still be significant depending on its structure. For example, in \cite{mao_soil_2018} a linear parameter-varying (LPV) model was identified for agro-hydrological dynamics approximation, where computational cost reduction was observed but still not sufficient for real-time online optimizations. 
	
	Model approximation based on neural network (NN) is a more generalized approach compared to traditional model identification. Motivated by the accuracy and computational efficiency of the NN model reduction observed in our previous work \cite{huang_comparative_nodate} and inspired by the structure of the LPV model developed in \cite{mao_soil_2018}, we propose a two-layer NN framework to approximate the agro-hydrological dynamics in this work. The proposed framework is shown to have better open-loop prediction performance compared to a benchmark single long short term memory (LSTM) NN. 
	
	The developed two-layer NN model is employed in a ZMPC controller in the presence of noise and weather disturbances. The control objective is to minimize the amount of irrigation required while keeping the soil moisture content inside the target zone. In order to address the plant-model-mismatch, open-loop bias correction is added to the model, while a shrinking target zone is employed for ZMPC. Two bias correction models are investigated with different updating frequencies and the one with better open-loop prediction performance is used in ZMPC. The shrinking ZMPC target zone is designed such that the shape of the zone can be tuned by modifying the hyper-parameters, of which the effects on the control performance are investigated.
	
	The paper is organized as follows. The agro-hydrological system of interest is introduced in Section \ref{pre}. Details of the proposed two-layer NN framework are discussed in Section \ref{model_design}, of which the open-loop validation is provided in Section \ref{V} with two plant-model-mismatch compensation strategies investigated in Section \ref{bias}. The ZMPC controller employed is introduced in Section \ref{controller}. The closed-loop simulation results are presented in Section \ref{results}. Finally, concluding remarks are provided in Section \ref{conclusion}. 
	
	\section{Preliminaries}	
	\label{pre}
\begin{figure}[!t]
		\centering
		\includegraphics[width=0.8\columnwidth]{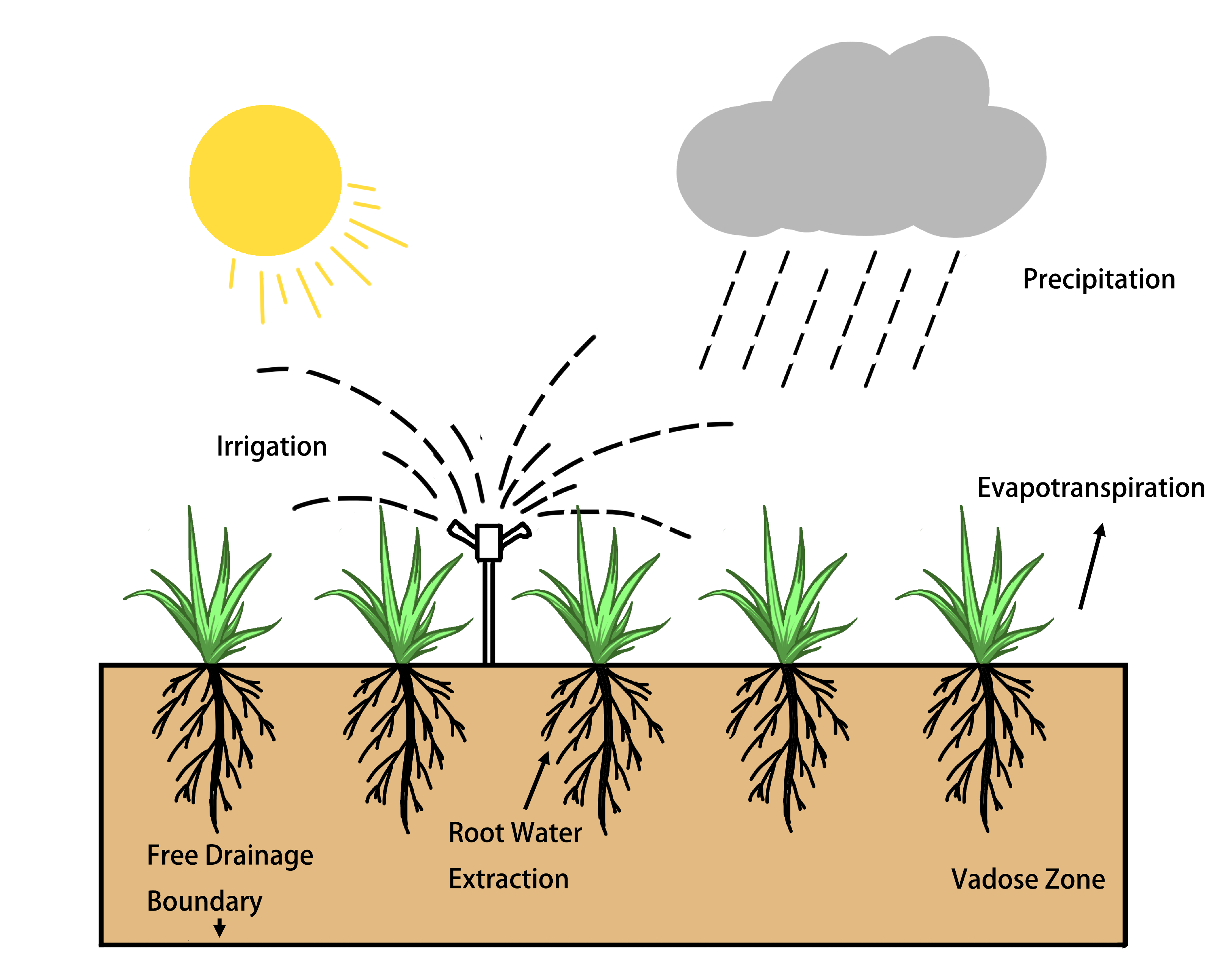}
		\caption{A schematic diagram of the considered agro-hydrological system.}
		\label{system}
	\end{figure}

	A schematic of the agro-hydrological system of interest is shown in Figure \ref{system} with grass being the crop of interest. The soil properties are assumed to be homogeneous horizontally and only the soil water dynamics on the vertical axis are considered. Irrigation is applied with a sprinkler and is assumed to reach the soil surface evenly. 
	Precipitation and evapotranspiration are considered as known disturbances based on weather forecasting data. It is assumed that precipitation reaches the soil surface in identical ways as irrigation. 
	%
	\subsection{Soil Water Dynamics}
	Based on first principles, the soil water dynamics can be modeled using the 1-dimensional Richards equation with the crop and weather information taken into account \cite{richards_capillary_1931}:
	\begin{equation}
		\label{Richards}
		c(h)\frac{\partial h}{\partial t} = \frac{\partial }{\partial z}\bigg[K(h)\bigg(\frac{\partial h}{\partial z} +1\bigg)\bigg] - \alpha(h)\frac{K_cET_0}{z_r}
	\end{equation}
	where $h$ $[m]$ is the capillary potential, $c$ $[m^{-1}]$ denotes the soil capillary capacity, and $K$ $[m/s]$ is the soil hydraulic conductivity. The $- \alpha(h)\frac{K_cET_0}{z_r}$ term captures the effect of weather and crop, where $\alpha(h)$ and $K_c$ are dimensionless factors, namely the water stress factor and the crop coefficient. $ET_0$ $[m]$ denotes the reference evapotranspiration and is assumed to be a known weather-dependent disturbance. $z_r =- 0.13$ m represents the rooting depth of the crop and is assumed to be time-invariant in this work. Note that $z$ $[m]$ represents the vertical axis of the soil with upwards as the positive direction. $c(h)$ and $K(h)$ are modeled by the van Genuchten-Mualem soil hydraulic model \cite{van_genuchten_closed-form_1980}: 
	\begin{equation}
		\label{capiliary_cap}
		c(h) = (\theta_s - \theta_r)\alpha n\bigg(1 - \frac{1}{n}\bigg)(-\alpha h)^{n-1}[1 + (-\alpha h)^n]^{\frac{1}{n} - 2}
	\end{equation} 
	\begin{equation}
		\label{hydraulic_cond}
			K(h) = K_s\Big[(1 + (-\alpha h)^n)^{-\big(1 - \frac{1}{n}\big)}\Big]^{\frac{1}{2}} 
			\Bigg[1 - \bigg[1 - \Big[(1 + (-\alpha h)^n)^{-\big(1 - \frac{1}{n}\big)}\Big]^{\frac{n}{n-1}}\bigg]^{1 - \frac{1}{n}}\Bigg]^2
	\end{equation}
	where $K_s$ $[m/s]$ denotes the saturated hydraulic conductivity, $\theta_s$ $[m^3/m^3]$ and $\theta_r$ $[m^3/m^3]$ represent the saturated and the residual soil moisture, respectively. $\alpha$ $[m^{-1}]$ and $n$ are the van Genuchten-Mualem parameters related to the soil properties. Sandy loam soil is considered in this work and the corresponding parameters are adopted from \cite{carsel_developing_1988} and are presented in Table \ref{soil_para}. 
	\begin{table}[!t]
		\small
		\centering
		\setlength{\arrayrulewidth}{0.5mm}
		\caption{Soil parameters of sandy loamy soil}	
		\renewcommand\arraystretch{1.2}
		\label{soil_para}
		\tabcolsep 20pt
		\begin{tabular}{lr}
			\hline
			$K_s \, [m/s]$ & 1.23e-5\\ 
			$ \theta_s \, [m^3/m^3]$& 0.41 \\
			$ \theta_r \, [m^3/m^3]$ & 0.065 \\
			$\alpha \, [m^{-1}]$ & 7.5\\
			$n$ & 1.89\\ 
			$m$ & 0.47\\ 
			\hline 
	\end{tabular}\end{table}

	\subsection{Boundary Conditions}
	The boundary conditions employed in this work are presented in (\ref{top_b})--(\ref{bottom_b}). The Neumann boundary condition is used at the soil surface as shown in (\ref{top_b}), while free drainage is assumed for the bottom boundary as described in  (\ref{bottom_b}). 
	\begin{equation}
		\label{top_b}
		\left.\frac{\partial h(t)}{\partial z}\right|_T = -1 - \frac{I(t) + P(t)}{K(h(t))}
	\end{equation}
	\begin{equation}
		\label{bottom_b}
		\left.\frac{\partial h(t)}{\partial z}\right|_B = 0
	\end{equation}
	where $I(t)$ $[m/s]$ and $P(t)$ $[m/s]$ denote the rate of irrigation and precipitation at time $t$. Note that interception caused by grass leaves is ignored in this work, only evapotranspiration, precipitation, and irrigation are considered.
	
	\subsection{Model Discretization}
	\label{Model_D}
	The system of equation (\ref{Richards}) is discretized over the vertical spatial axis following the approach proposed in \cite{richards_capillary_1931}, which leads to an ODE system: 
	\begin{equation}
		\label{sys_o}
		\dot{x}(t)=f(x(t),u(t))
	\end{equation}
	where $x\in\mathbb{R}^{26}$ and $u\in\mathbb{R}^{1}$ are the state and input vectors respectively, while $f$ defines the nonlinear system dynamics. The total depth of soil considered is $0.5$ m and is evenly discretized into 26 nodes. The water dynamics at the center of each node are used to represent the dynamics of the entire node. In this work, irrigation rate $I(t)$ $[m/s]$ is the system input, and $h$ $[m]$ at the discretized nodes represents the system states. The system output $y\in\mathbb{R}^{1}$ is the volumetric water content $\theta$ $[m^3/m^3]$ at the rooting depth $z_r$, which is an algebraic function of the state vector defined by the soil-water retention equation of van Genuchten (\ref{soil_mois}). Eqn. (\ref{sys_o}) and Eqn. (\ref{soil_mois}) are used to generate all training datasets. 
	\begin{equation}
		\label{soil_mois}
		\theta(h) = (\theta_s - \theta_r)\bigg[\frac{1}{1 + (-\alpha h)^n}\bigg]^{1 - \frac{1}{n}} + \theta_r
	\end{equation}  
	\section{Proposed two-layer neural network}
	\label{model_design}
	To reduce the computational complexity of the first-principle model, a two-layer NN framework is proposed to approximate its nonlinear dynamics. In this section, details regarding the proposed framework will be discussed. Section \ref{3.1} introduces the motivation and provides an overview of the proposed framework. Sections \ref{DG_1} and \ref{DG_2} discuss the data generation procedure for the individual layers of the framework respectively. 
	\subsection{Overview}
	\label{3.1}
	\begin{figure}[!t]
		\centering
		\includegraphics[width=1\columnwidth]{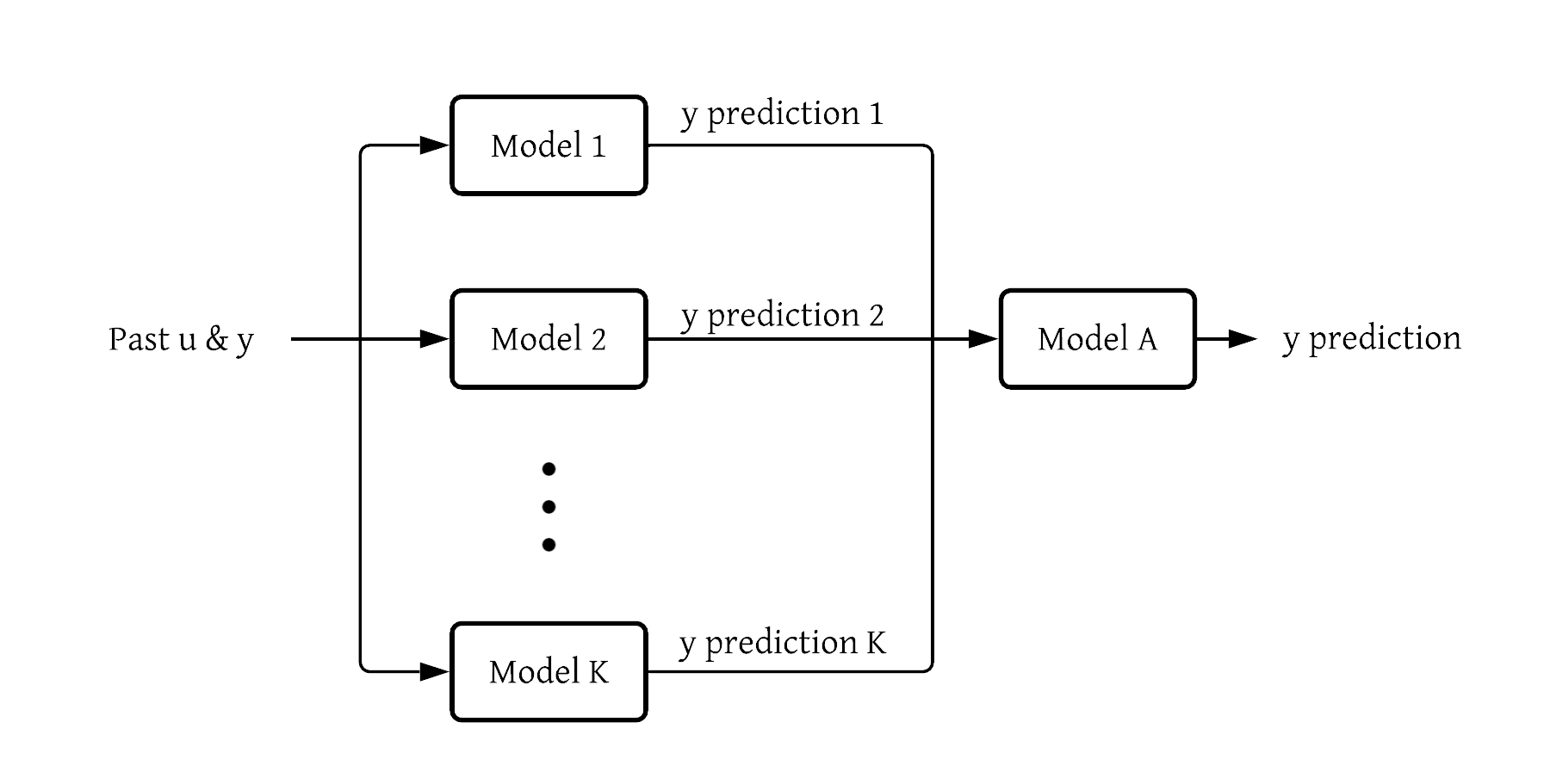}
		\caption{A schematic diagram of the proposed two-layer NN framework.}
		\label{2_layer_NN}
	\end{figure}
	
	 In this work, LSTM NNs will be used to approximate the original nonlinear system dynamics due to their ability of efficiently handle sequential data \cite{6795963,  lipton_critical_2015}. The motivation for designing the proposed framework is that a single LSTM is not able to capture the complex nonlinear dynamics of the agro-hydrological system over the entire operating range of interest. It was noticed, however, within a smaller operating region, a single LSTM provides satisfying modeling performance. Detailed simulation results will be presented in Section \ref{V}. 
	
	A schematic diagram of the proposed framework is presented in Figure \ref{2_layer_NN}. The first layer contains $K$ sub-models, namely Model $1$-- Model $K$, which will be referred to as $M_1, M_2, \cdots, M_K$ in the following text. The second layer consists of a single model (Model A, referred to as $M_A$). Each sub-model in the first layer is responsible for a sub-operating region. The union set of the sub-operating regions forms the entire operating region of interest. $M_1-M_K$ are LSTMs while $M_A$ is a fully connected NN. At the current time step $t$, $M_1-M_K$ take the system input $u$ and output $y$ obtained from $p$ previous time steps as NN inputs:
	\[
	NN_{in}(t) = \left[\begin{array}{cc}
		u(t-p+1), y(t-p+1) \\
		u(t-p+2), y(t-p+2) \\
		\cdots\\
		u(t), y(t) \\\end{array}\right], 
	\] 
	and returns $K$ one-step-ahead predictions regarding the system output:
	\[NN_{out_1} = [\bar{y}_1(t+1), \cdots, \bar{y}_K(t+1)]^T\]
	Recall that each sub-model is only reliable in a sub-operating region. Inspired by Mao et al. \cite{mao_soil_2018}, the proposed framework adopts the idea of the LPV models. The model performance can be improved by combining predictions made by multiple sub-models. The one-step-ahead predictions made by $M_1 - M_K$ are fed to $M_A$, where the final prediction of the system output at the next time step ($NN_{out_2} = y_{pred}(t+1)$) is obtained. 
	
	Overall, at any time step $t$, the proposed framework takes $NN_{in}(t)$ as input and returns the one-step-ahead prediction of the system output $y(t+1)$, which can be expressed as a nonlinear function:
	\begin{equation}
		\hat{y}(t+1|t)=g(NN_{in}(t))
	\end{equation}
	
	\subsection{Design and training of first-layer NNs}

	\label{DG_1}
	\begin{table}[!t]
		\small
		\centering
		\caption{Operating ranges of the sub-models}	
		\renewcommand\arraystretch{2}
		\label{sub_models}
		\tabcolsep 10pt
		
		\begin{tabular*}{0.5\textwidth}{ccc}\hline
			& Operating range & \# of LSTM layers\\ \hline
			$M_1$ & $y \in [0.12, 0.27]$ & 1\\
			$M_2$ & $y \in [0.21, 0.32]$ & 1\\ 
			$M_3$ & $y \in [0.29, 0.40]$  & 2\\ \hline
		\end{tabular*}
	\end{table}

	\begin{figure}
		\centering
		\includegraphics[width=0.9\columnwidth]{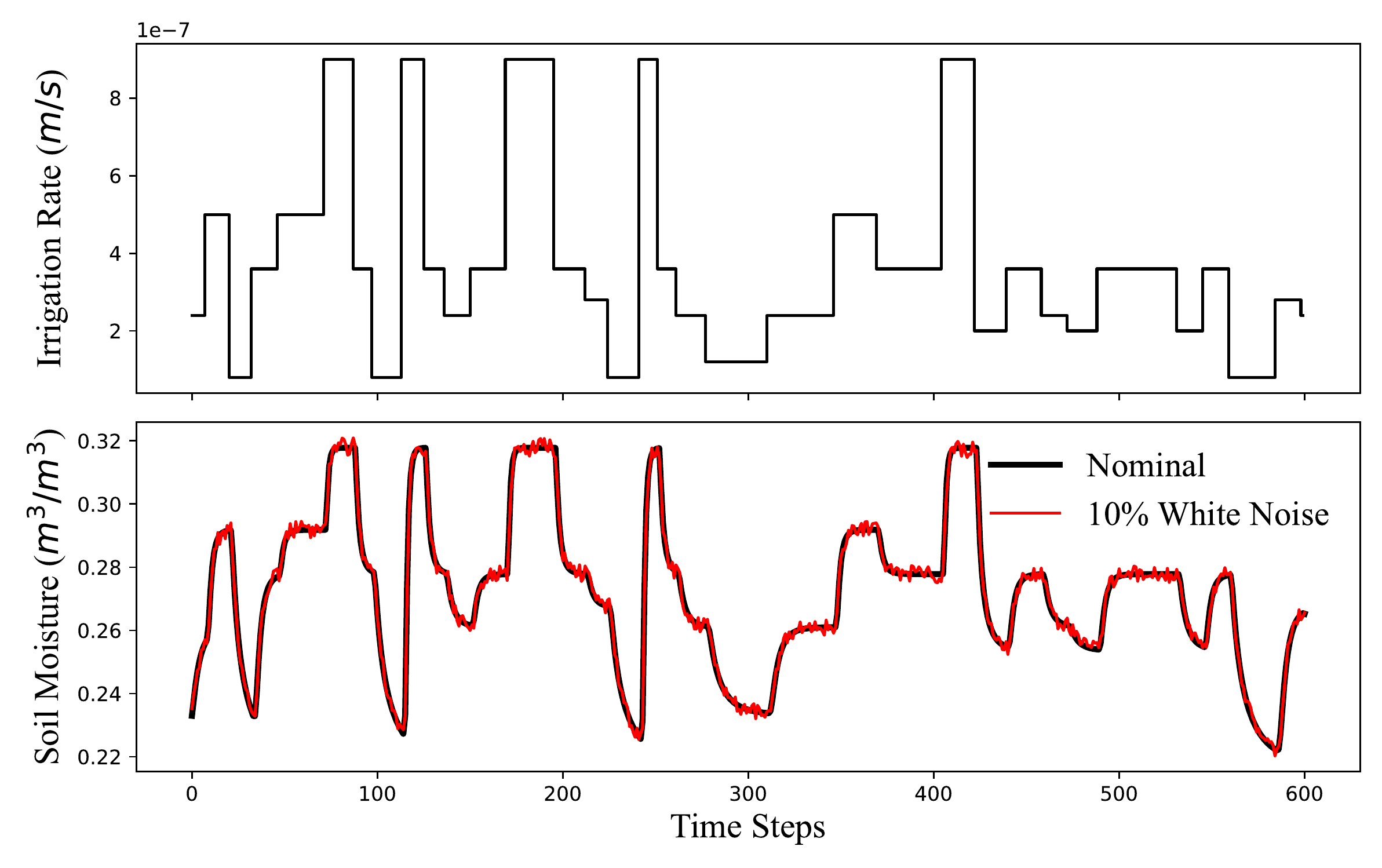}
		\caption{A segment of the training dataset of $M_2$}
		\label{sub}
	\end{figure}
	
	As discussed in the previous section, the first layer of the proposed framework consists of $K$ LSTMs. In this work, $K = 3$ is used. 
	The operating regions in terms of the system output for each sub-model are summarized in Table \ref{sub_models}. The number of LSTM layers employed in the models is also included. Each model consists of 1 or 2 LSTM layers, followed by a fully connected layer. Activation functions \texttt{`sigmoid'} and \texttt{`tanh'} are used. Both activation functions have smooth dynamics, where \texttt{`tanh'} function provides steeper responses compare to \texttt{`sigmoid'} function. Each sub-model is equivalent to a nonlinear function:
	\[m_i:\mathbb{R}^{20 \times 2}\rightarrow\mathbb{R}^{1}, \quad i = 1, 2, 3\] 
	where the model input is a matrix with $20$ rows and $2$ columns, and the output is a scalar. $20$ is the number of the past input-output data points ($p = 20$), while $2$ is the number of system input ($n = 1$) plus the number of system output($l = 1$). All models have a sampling time of 2 hours.  
	
	Three training datasets that vary in the corresponding operating region are generated for the sub-models respectively. Each dataset contains 30000 data points. Multi-level pseudorandom input signals are employed to generate all three datasets, which are fed into the discrete ODE system (\ref{sys_o}) and the algebraic equation (\ref{soil_mois}) for open-loop simulations. A multi-level pseudorandom signal is similar to a pseudorandom binary signal, except that it has multiple levels, which helps to stimulate the nonlinear dynamics of the agro-hydrological system. More details regarding the design of multi-level pseudorandom inputs can be found in our previous work \cite{huang_comparative_nodate}. Figure \ref{sub} presents a segment of the training input signal and the corresponding system output for $M_2$. Note that a random noise signal $\epsilon$ is added to the output trajectory to mimic real process operations, which is shown in Figure \ref{sub}(b) in red. The max magnitude of $\epsilon$ is defined as follows:
	\[\epsilon_{max} = 0.1[y_{max} - y_{min}]\]
	where $y_{max}$ and $y_{min}$ are the maximum and minimum value of the system output in the dataset. 
	
	All datasets need to be reorganized to match the input-output structure of the NNs. Scaling is also required such that the magnitudes of the scaled data are less than 1. All simulations are carried out in Python. TensorFlow \cite{tensorflow2015-whitepaper} is used to train the NN models. Readers may refer to our previous work \cite{huang_comparative_nodate} for more details regarding data pre-processing and NN training. 
	
	\subsection{Design and training of second-layer NN}
	\label{DG_2}
	
	\begin{figure}
		\centering
		\includegraphics[width=0.9\columnwidth]{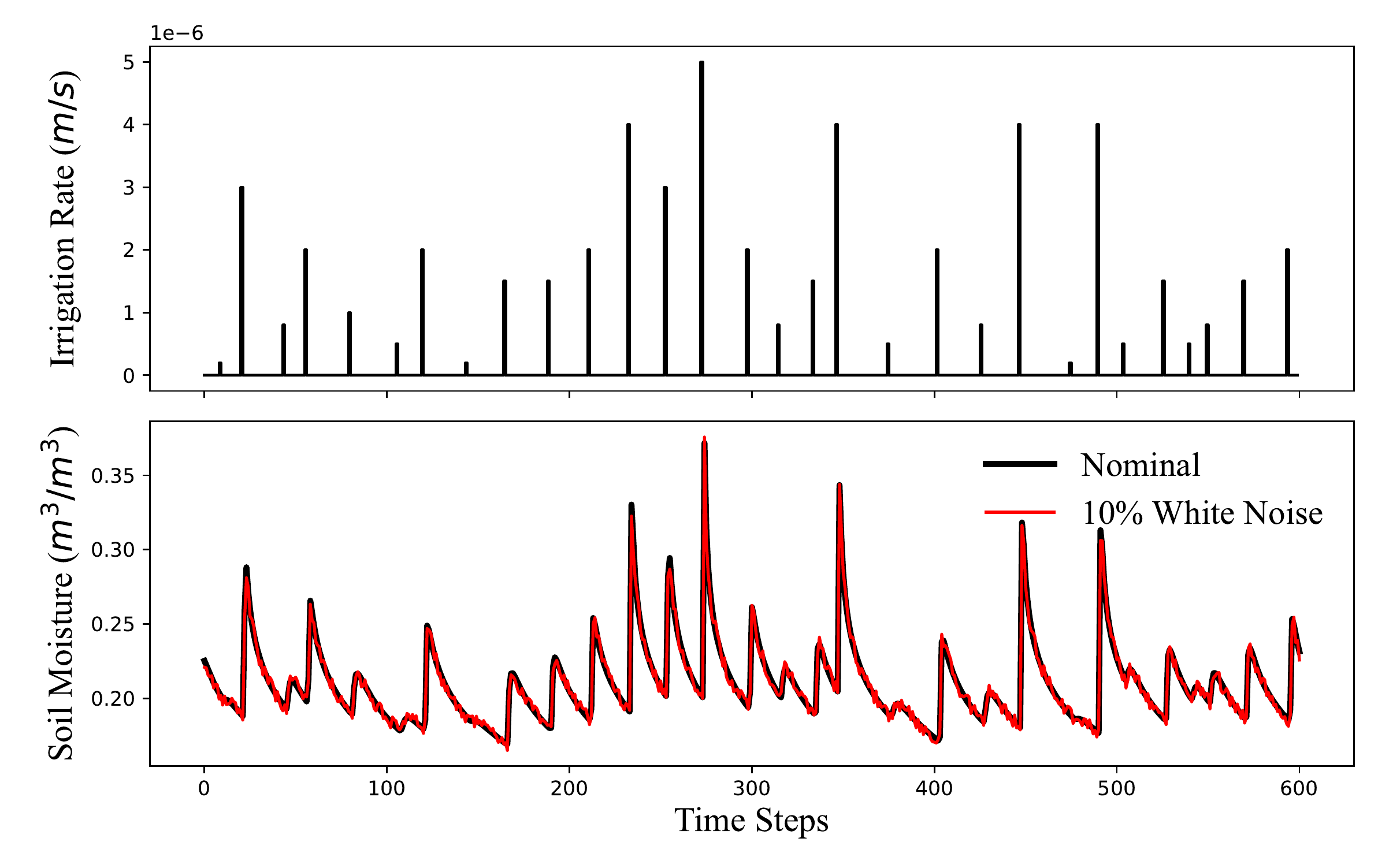}
		\caption{A segment of the training dataset of $M_A$}
		\label{top}
	\end{figure}
	
	The second layer of the proposed framework consists of a fully connected NN equivalent to the following nonlinear function:
	\[m_A:\mathbb{R}^{3}\rightarrow\mathbb{R}^{1}\]  
	The inputs to model $M_A$ are the one-step-ahead predictions of the system output made by $M_1$, $M_2$ and $M_3$, while the output is the actual system output prediction. Two fully connected layers with the activation function \texttt{`sigmoid'} are used in $M_A$. 
	
	Under the objective of predicting the system output over the entire operating range of interest, the training dataset of $M_A$ covers the entire operating range. A section of the training dataset including 100000 points is presented in Figure \ref{top}, where the input signal and the corresponding output trajectories without and with white noise added are presented. Note that the training dataset of $M_A$ spans all the sub-operating regions with more aggressive dynamics. Instead of continuous irrigation with lower magnitudes, an impulse irrigation signal is used to provide a more aggressive output response that varies in a larger operating range, which is desired for $M_A$ training. 
	
	Similar to that discussed in the previous section, reconstruction and scaling are necessary for this dataset. Note that the inputs to $M_A$ are not the system input and output. The dataset is passed through the pre-identified sub-models after the data pre-processing. The predictions made by the sub-models are collected, which are the designed inputs for $M_A$ and are used to train $M_A$.

	\section{Model Validation}
	\label{V}
	The prediction performance of the proposed framework is validated in this section. The maximum number of prediction steps is $N = 20$ for this work, which is also the control horizon of interest for following ZMPC applications. Section \ref{Vali} presents the validation results of the sub-models. The proposed framework is then compared with a single LSTM in Section \ref{Vali_o}. 
	\subsection{Validation of the first-layer sub-models}
	\label{Vali}
	\begin{table}[!t]
		\small
		\centering
		\caption{NRMSE of multi-step-ahead-predictions of the sub-models}	
		\renewcommand\arraystretch{1.2}
		\label{OL_S}
		\tabcolsep 10pt
		
		\begin{tabular*}{0.5\columnwidth}{cccc}\hline
			\# of prediction steps & $M_1$ & $M_2$ & $M_3$\\ \hline
			1 & 0.015 & 0.016 & 0.073\\
			10 & 0.042 & 0.049 & 0.111\\ 
			20 & 0.060 & 0.050 & 0.111\\ \hline
		\end{tabular*}
	\end{table}	
	
	The multi-step-ahead prediction performances of $M_1 - M_3$ are presented in this section. Validation datasets are obtained in similar manners as the training datasets with 20\% white noise added to the system output. This helps to better test the prediction performance of the models. At any given time step, noise-treated validation data are fed to the model for one-step-ahead prediction. Starting from the second time instant, the prediction made by the model at the previous time instant will be used as the initial condition of the current time instant. This procedure is repeated 20 times, providing twenty-steps-ahead predictions. 
	
\begin{figure}
		\centering
		\includegraphics[width=0.8\columnwidth]{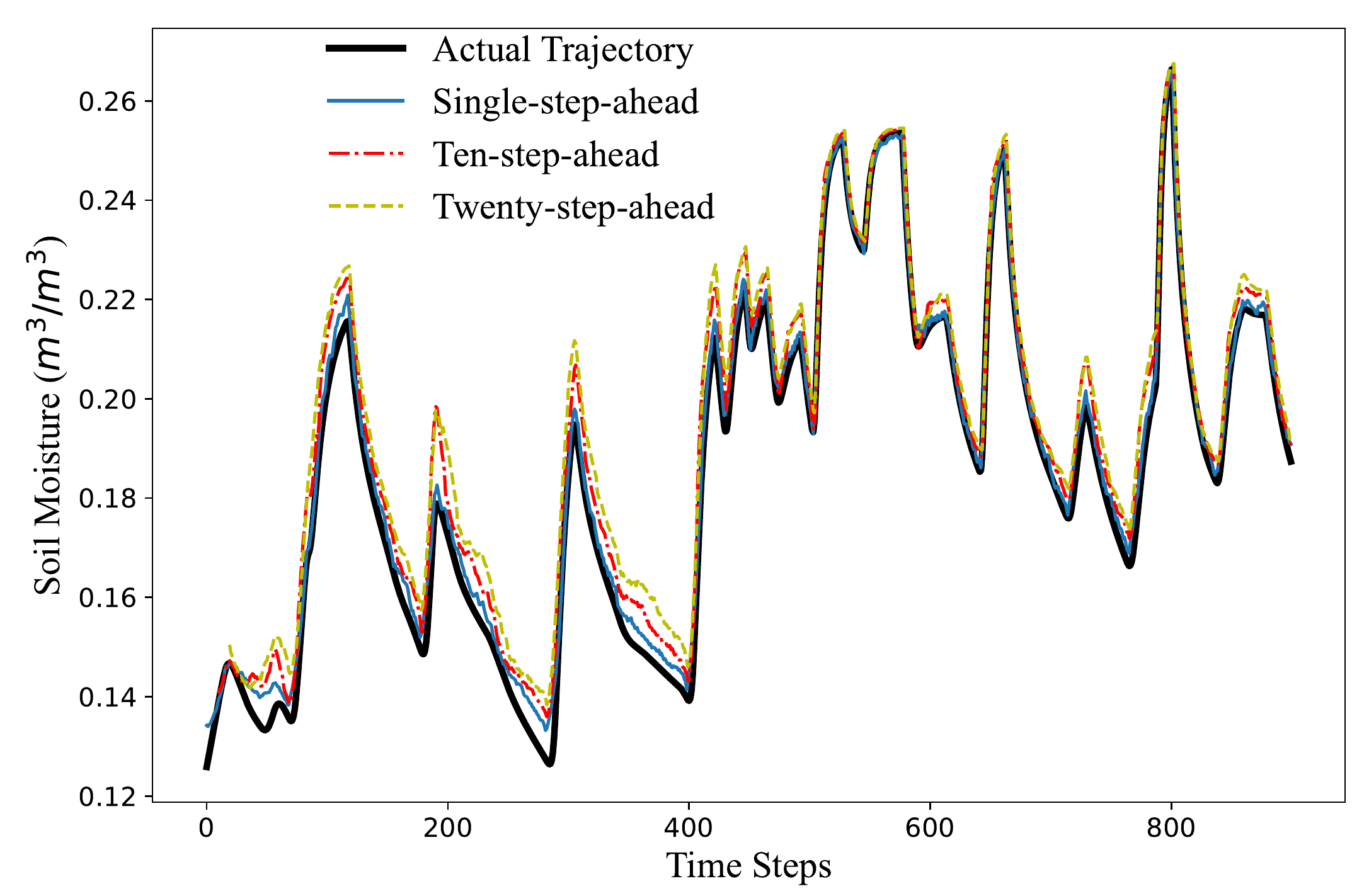}
		\caption{One-step-ahead prediction (solid), ten-step-ahead prediction (dash-dotted), and twenty-step-ahead prediction (dashed) of $M_1$ compared to the actual trajectory (dark solid).}
		\label{SM_1}
	\end{figure}

\begin{figure}
		\centering
		\includegraphics[width=0.8\columnwidth]{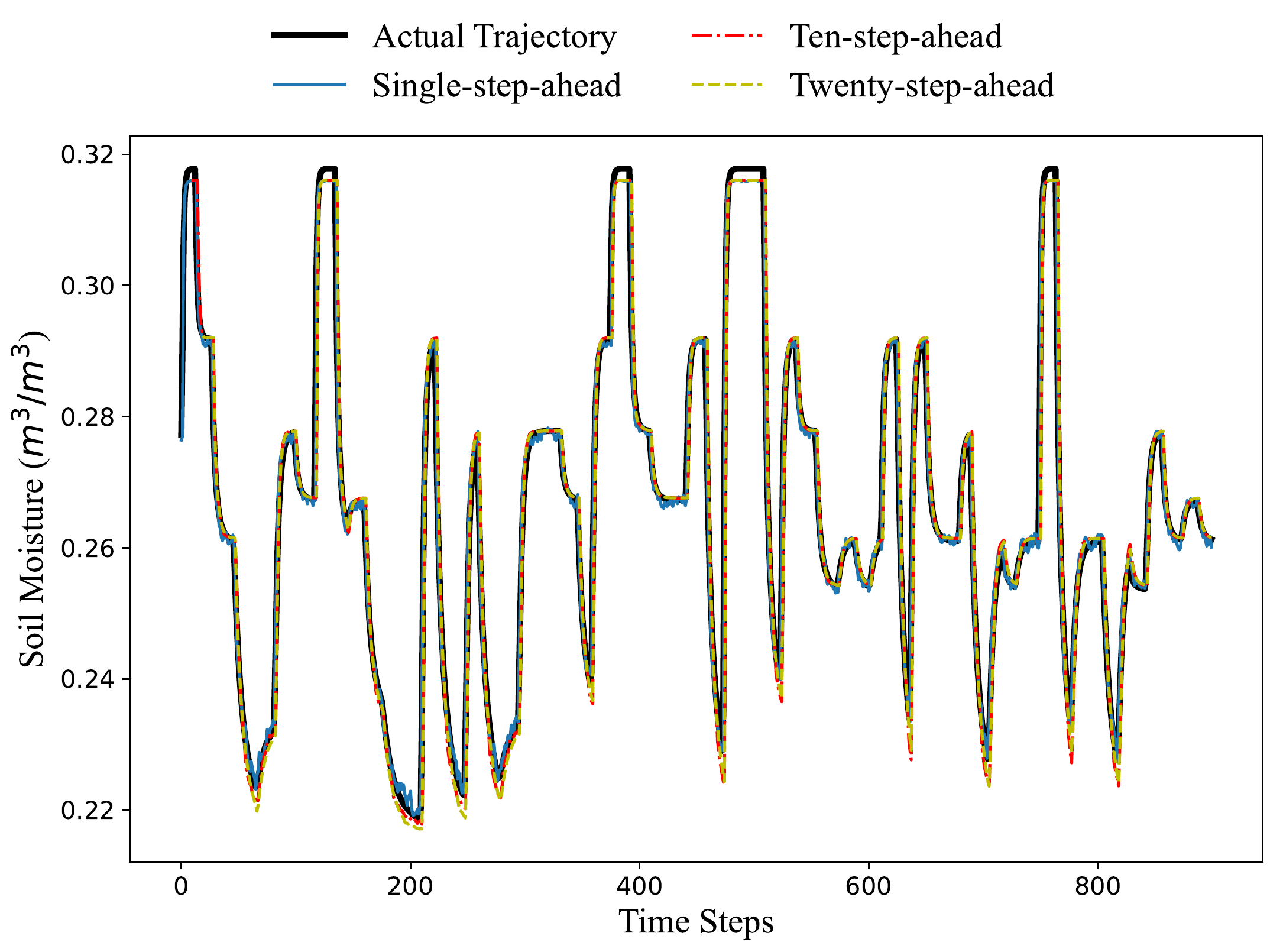}
		\caption{Multi-step-ahead prediction performance of $M_2$.}
		\label{SM_2}
	\end{figure}

\begin{figure}
		\centering
		\includegraphics[width=0.8\columnwidth]{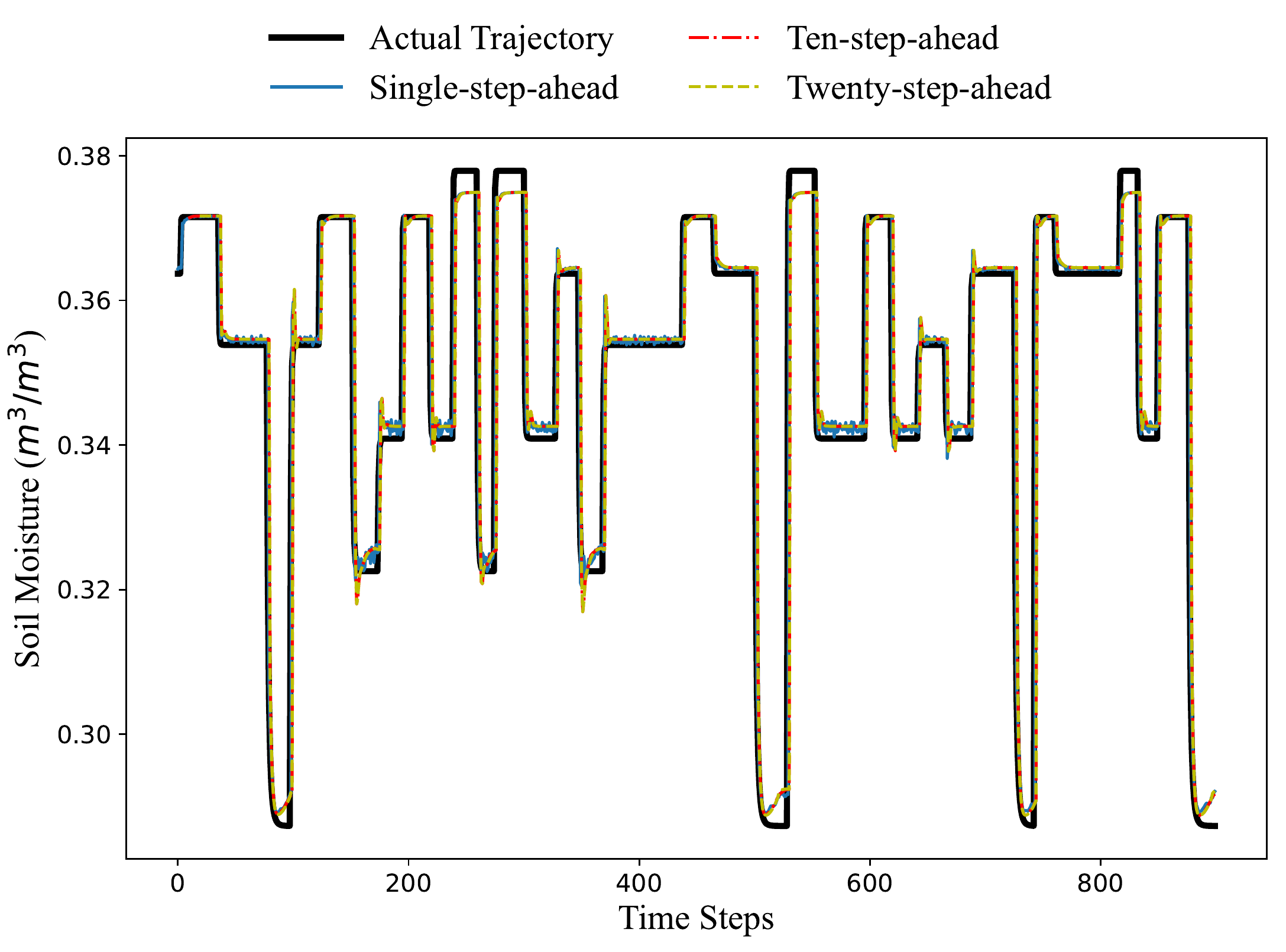}
		\caption{Multi-step-ahead prediction performance of $M_3$.}
		\label{SM_3}
	\end{figure}

	The prediction performance is measured by the normalized root mean square error (NRMSE) of the multi-step-ahead predictions in the desired operating range of each model and is summarized in Table \ref{OL_S}. The NRMSE is defined as follows:
	\begin{equation}
		NRMSE = (y_{max} - y_{min})^{-1}\sqrt{\frac{\sum_{i=1}^{t_f} (y_{act}(i) - y_{pred}(i))^2}{t_f}}
	\end{equation}
	where $t_f$ is the total number of validation data points, $y_{act}$ and $y_{pred}$ are the true and predicted system outputs, respectively. Figures \ref{SM_1}, \ref{SM_2}, and \ref{SM_3} present the multi-step-ahead open-loop prediction performance of $M_1$, $M_2$, and $M_3$ respectively. The same line styles are used in Figures \ref{SM_1}--\ref{SM_3} for consistency. 
	
	All three models provide reasonable multi-step-ahead prediction without delays in the presence of noise. When the magnitude of the system output is low, small prediction offsets can be observed for $M_1$ in Figure \ref{SM_1}. Figure \ref{SM_2} shows only some minor offsets exist at the local extremes for $M_2$. For both $M_1$ and $M_2$, the effect of noise seems to be minor. For $M_3$, Figure \ref{SM_3} indicates that the noise has a stronger impact on the single-step-ahead prediction, which deviates and becomes insignificant as the number of prediction steps increases. It is also noticed that when the output converges to steady-states, the impact of the noise is the most significant. Similar to $M_1$ and $M_2$, $M_3$ also has some minor offsets at the local extremes.
	
	For different operating ranges, different system output dynamics are observed. When the soil moisture content and the irrigation rate are low, no obvious steady-state is observed (Figure \ref{SM_1}). As the operating range shifts to the higher magnitude side, the soil moisture content tends to saturate to steady-states under continuous irrigation (Figure \ref{SM_3}). 
	
	\subsection{Validation of the proposed two-layer framework}
	\label{Vali_o}
	\begin{figure}
		\centering
		\includegraphics[width=0.9\columnwidth]{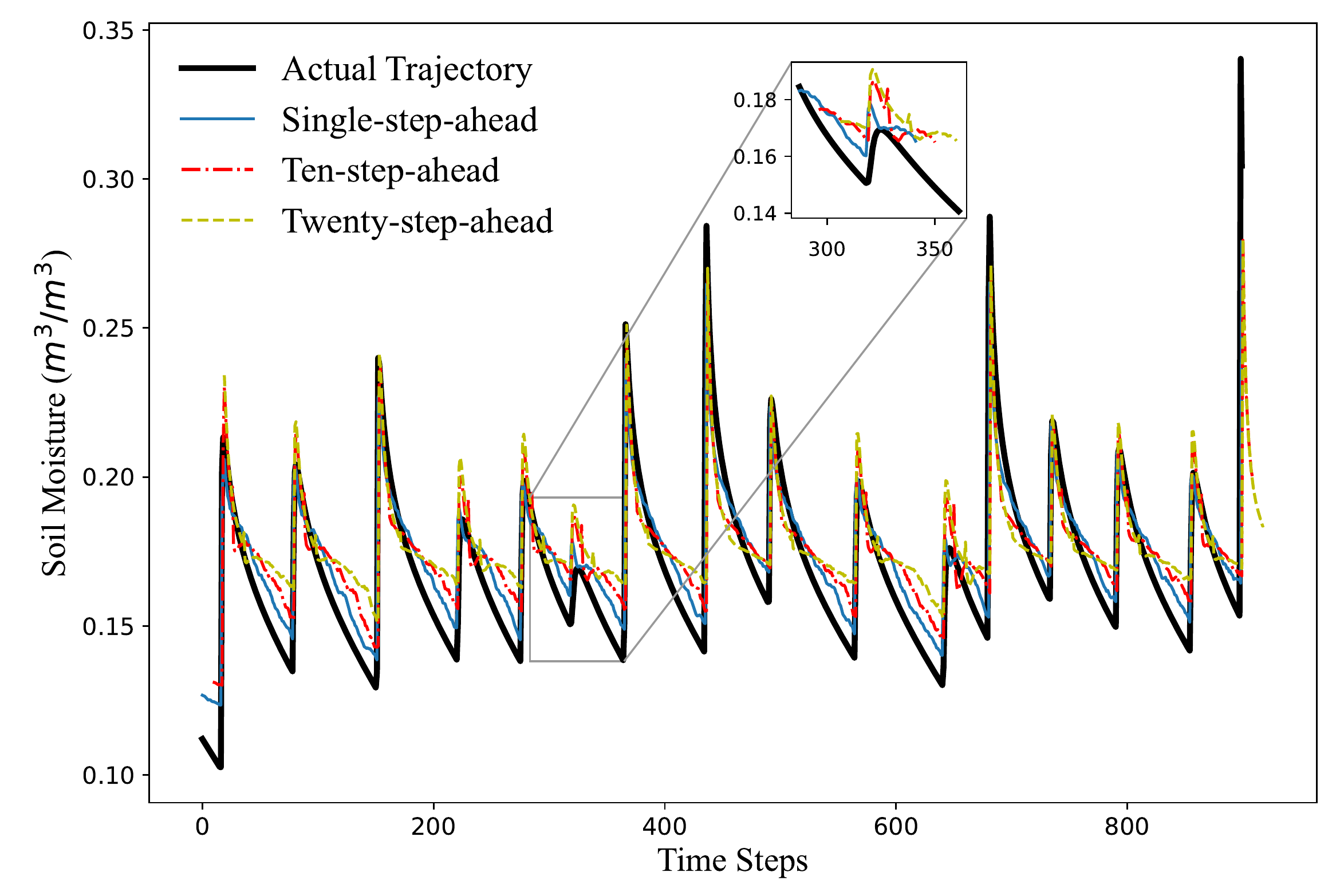}
		\caption{Multi-step-ahead prediction performance of an single LSTM.}
		\label{Multi_step_1_layer}
	\end{figure}
	
	\begin{figure}
		\centering
		\includegraphics[width=0.9\columnwidth]{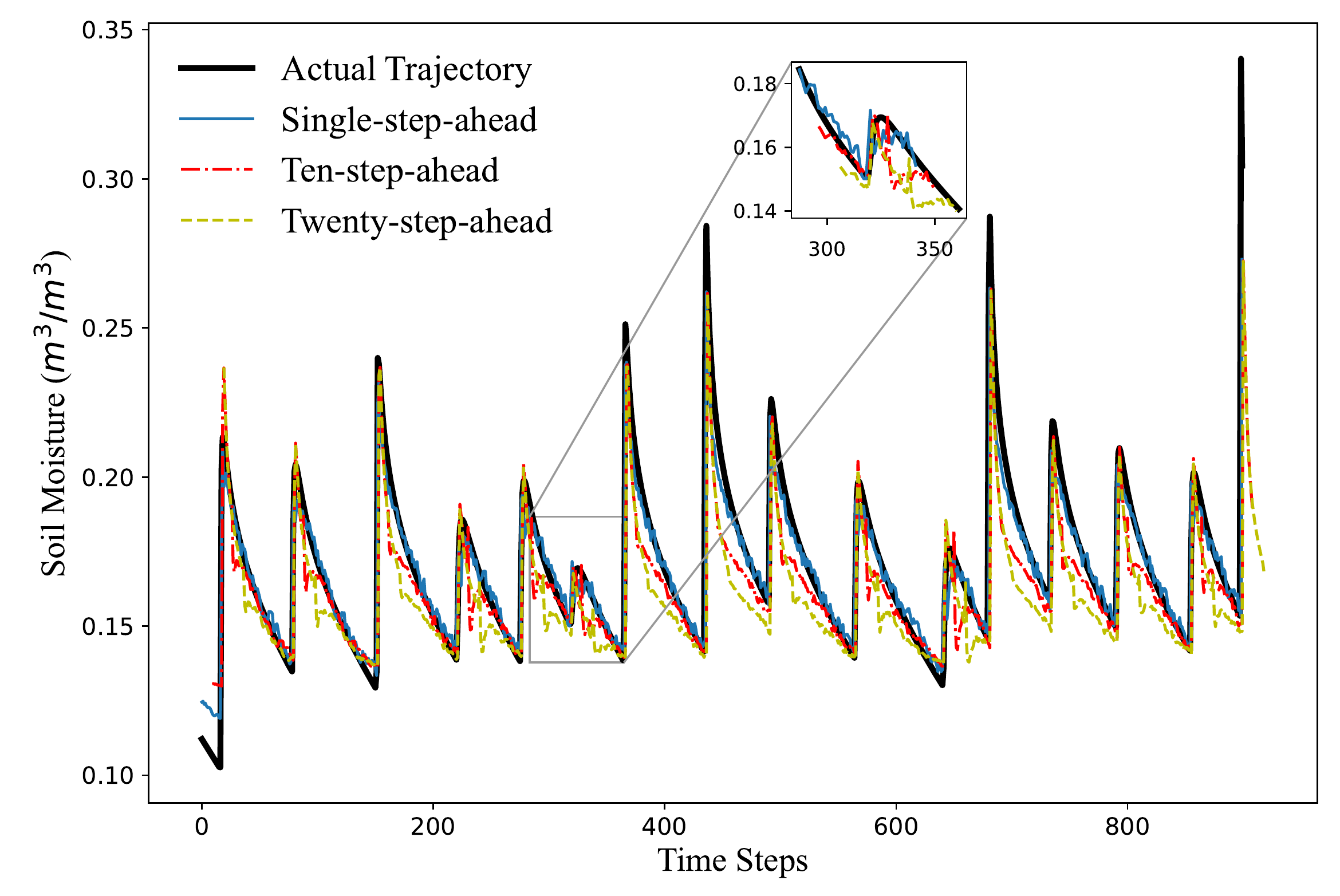}
		\caption{Multi-step-ahead prediction performance of the two-layer NN framework.}
		\label{Multi_step_2_layer}
	\end{figure}
	
	The multi-step-ahead prediction performance of the proposed framework is investigated in this section. A single LSTM that aims to predict the agro-hydrological dynamics over the entire operating region is used as the benchmark approximation model. The benchmark LSTM has the same input ($NN_{in}$) and output ($NN_{out_2}$) structures as the proposed two-layer NN framework, which are discussed in Section \ref{3.1}. The single LSTM has 2 LSTM layers followed by a fully connected output layer. Both the LSTM layers use the activation function \texttt{`sigmoid'} while the output layer uses the activation function \texttt{`tanh'}. Note that the dataset employed for training the benchmark LSTM is the same as that used to train the two-layer NN model. 
	
	The same validation dataset is used to compare the performance of the two approximation approaches, which spans the entire operating range of interest. The open-loop prediction trajectories of the system output based on the benchmark LSTM and the proposed framework are presented in Figures \ref{Multi_step_1_layer} and \ref{Multi_step_2_layer} respectively. The single-step-ahead, ten-step-ahead, and twenty-step-ahead predictions are presented in each figure with the same line styles as those employed in the previous section. The NRMSE values under the two cases and the percentage difference $\triangle$ between them are summarized in Table \ref{OL}. The percentage difference is calculated with respect to the NRMSE of the benchmark LSTM:
	\[\triangle = \frac{NRMSE_{LSTM} - NRMSE_{proposed}}{NRMSE_{LSTM}}\]
	
	As presented in Figure \ref{Multi_step_1_layer}, the LSTM can respond to aggressive dynamics efficiently without delays. However, significant offsets are observed at the local extremes with larger magnitudes. A section of the trajectories is amplified, which can be seen on the top-right corner of Figure \ref{Multi_step_1_layer}. At both the local maximum and the minimum, significant offsets can be observed in the zoomed-in plot, of which the predicted local minimums are at the level of the true local maximum. 
	Better prediction performance is observed in Figure \ref{Multi_step_2_layer}. To be specific, the dynamics at the local minimums can be captured by the proposed two-layer framework much better compared to the LSTM. From the zoomed-in window on the top right, it can be observed that the predictions of the local minimum and maximum are more accurate compared to the LSTM. Table \ref{OL} indicates comparison results, showing that the NRMSE values of the proposed framework are significantly lower than those obtained with the LSTM for all different prediction steps. The improvement of the proposed framework is the most significant with one-step-ahead predictions and slowly reduces as the prediction steps increase.
	\begin{table}[!t]
		\small
		\centering
		\caption{NRMSE of multi-step-ahead-predictions of the NN models}	
		\renewcommand\arraystretch{1.2}
		\label{OL}
		\tabcolsep 5pt
		\begin{tabular*}{0.56\columnwidth}{cccc}\hline
			\# of Pred. steps & Single LSTM& two-layer NN & \% Diff. $\triangle$\\ \hline
			1 & 0.076 & 0.044 & 42.5\\
			10 & 0.096 & 0.082 & 14.5\\ 
			20 & 0.100  & 0.093 & 7.14\\ \hline
		\end{tabular*}
	\end{table}	
	
	It is noticed that the proposed framework cannot accurately approximate the local maximums with larger magnitudes. To account for this problem, we introduce a bias-and-scale-updated model to capture the plant-model-mismatch, of which the details are provided in the following section.

	\section{Plant-Model-Mismatch Correction}
	\label{bias}
	To capture the plant-model-mismatch, two correction approaches are investigated based on the proposed two-layer NN framework. The first approach is to add a simple bias term $b_1$ based on the prediction error:
	\begin{equation}
		y_{act} = y_{pred} + b_1
	\end{equation}
	where $y_{act}$ and $y_{pred}$ represent the actual output measurement and the prediction made by the two-layer NN framework respectively.  
	
	The second approach is to fit a first order linear model that maps the predicted output to the actual measurement:
	\begin{equation}
		y_{act} = a\cdot y_{pred} + b_2
	\end{equation}
	where $a$ and $b_2$ are parameters to be identified. 
	
	For both approaches, an updating frequency $f$ of the parameters needs to be determined first. In this work, we are interested in accurately predicting the system dynamics 20 steps into the future, thus the updating frequency should be a divisor of 20. To test the performance of the two approaches, the validation dataset used in the previous section is employed. The updating policies of the two approaches are presented in Algorithms \ref{alg1} and \ref{alg2}, respectively. In both algorithms, $t$ represents the current time step, $i$ denotes the step of predictions made ahead of time $t$, and $\eta(i)$ represents the prediction error at step $i$.
	
	At any given time instant $t$, the parameters are first initialized with some initial guess. The initial condition $NN_{in}(t)$ is provided to the proposed framework for multi-step ahead predictions. In the next $N = 20$ steps, predictions will be made with the correction model considered, which will be updated every $f$ time steps for $\frac{N}{f}$ times. 
	
	The updating policy of the first approach is to take the average of the accumulated prediction errors over the last $f$ prediction steps. For the second approach, a constrained optimization problem that finds the best-fit parameters for the linear model is solved. Note that the parameters are held in bounded compact sets, which help to limit the parameters in a reasonable range. To be more specific, $a \in [0.8, 1.5]$ and $b_2 \in [-0.2, 0.3]$ are used. If a simple linear-regression function is employed, the resulting parameters might be impractical. The toolbox CasADi \cite{Andersson2018} is employed to solve the optimization problems in Python. 
	
	\begin{algorithm}
		\caption{Parameter updating algorithm for the first correction approach}\label{alg1}
		\begin{algorithmic}
			\State Set the number of prediction steps $N = 20$
			\State Set the updating frequency $f = f$
			\State Initialize $b_1 = 0$
			\State At each time step of the dataset t, set $i = 1$
			\State Provide the model with the initial condition $y_{pred}(0) = y_{act}(t)$
			\While{$i \leq N$}
			\State Predict the future system output $y_{pred}(i) = NN_{out_2}(i) + b_1$
			\State Record the error with respect to the actual output value $\eta(i) \gets (y_{act}(t+i) - y_{pred}(i))$
			\If{$f$ is a divisor of $i$}
			\State Update $b_1 = \frac{1}{f}\sum_{j = i-f+1}^i \eta(j)$ 
			\EndIf
			\State Set $i = i+1$
			\EndWhile

		\end{algorithmic}
	\end{algorithm}

	\begin{algorithm}
	\caption{Parameter updating algorithm for the second correction approach}\label{alg2}
	\begin{algorithmic}
		\Ensure $a \in \mathbb{A}$, $b_2 \in \mathbb{B}$
		\State Set the number of prediction steps $N = 20$
		\State Set the updating frequency  $f = f$
		\State Initialize $a = 1$, $b_2 = 0$ 
		\State At each time step of the dataset t, set $i = 1$
		\State Provide the model with the initial condition $y_{pred}(0) = y_{act}(t)$
		\While{$i \leq N$}
		\State Predict the future system output $y_{pred}(i) = a \cdot NN_{out_2}(j) + b_2$
		\If{$f$ is a divisor of $i$}
		\State Update $a$ and $b_2$ by solving the optimization problem 
		\State $a^*, b_2^* = \min_{a, b_2} \; \sum_{j = i-f+1}^i (y_{act}(t+j) - y_{pred}(i))$ 
		\EndIf
		\EndWhile
	\end{algorithmic}
	\end{algorithm}

	Recall the prediction horizon of interest $N = 20$. Thus updating frequencies $f = 1, 2, 5, 10$ are tested and compared. The correction performance is measured by the average absolute prediction error $\upsilon_{mm}$:
	\[\upsilon_{mm} = \frac{1}{N \cdot t_f}\sum^{t_f}\sum^N|y_{act} - \hat{y}_{pred}|\]
	where the same definitions are kept for all variables. The absolute prediction errors of the two correction approaches are summarized in Table \ref{bias_comp}. The absolute prediction error of just the proposed NN framework is as well included in Table \ref{bias_comp} and is used as the comparison benchmark.
	
	\begin{table}[!t]
		\small
		\centering
		\caption{Performance of the mismatch correction approach in terms of average absolute error}	
		\renewcommand\arraystretch{1.2}
		\label{bias_comp}
		\tabcolsep 8pt
		
		\begin{tabular*}{0.45\columnwidth}{cccc}\hline
			$f$ & No corr.  & Single bias & Linear Model \\ \hline
			1 & 0.101 & 0.066 & -\\
			2 & 0.101 & 0.064 & 0.063\\ 
			5 & 0.101 & 0.195 & 0.090\\ 
			10 & 0.101 & 0.265 & 0.098 \\ \hline
		\end{tabular*}
	\end{table}	
	
	With $f = 5, 10$, the single-bias correction approach leads to more significant errors, which is not applicable. The linear model correction shows very minor performance improvements. When $f = 2$, similar improvements in the prediction performance are observed in both approaches. The single bias correction performs similarly for $f = 1$ and $f=2$. For the linear model correction, $f = 1$ is not applicable. Considering the computational complexity of the two approaches, the single-bias correction with $f = 2$ is selected.
	
	\section{Controller Design}
	\label{controller}
	A ZMPC controller with a shrinking target zone is employed in this work and is presented in this section. 
	The major target here is to provide sufficient irrigation such that the crop can live and grow healthily, which implies the soil moisture content at the crop root zone needs to be kept in a bounded range instead of one particular point. Compared to a conventional tracking MPC, ZMPC provides more degrees of freedom to the system and can handle economic objectives better, which fits the objective of this work. 
	
	In the basic design of a ZMPC controller, the target zone is kept time-invariant over the control horizon. In this work, we propose a ZMPC design with a time-variant target zone that shrinks over the control horizon to handle plant-model-mismatch.
	The proposed ZMPC formulation is presented as follows:
	\begin{subequations}
		\label{ZEMPC}
		\begin{empheq}{align}
			\label{ZEMPC_1}
			\min_{u(t_j|t_i)\in S(\triangle)} \; &\sum_{j = i}^{i+N-1} || \hat{y}(t_j|t_i) - y_z(t_j|t_i) ||^2_Q + || u(t_j|t_i)||^2_R  \\
			\label{ZEMPC_2}
			\textmd{s.t.}\ :\; &\hat{y}(t_{j+1}|t_i)=g(NN_{in}(t_j|t_i)), \quad j = i, i+1,\cdots, i+N-1 \\
			\label{ZEMPC_3}
			&\hat{y}(t_i|t_i) = y(t_i) \\
			\label{ZEMPC_4}
			&u(t_j|t_i) \in \mathbb{U}, \quad j = i, i+1,\cdots, i+N-1 \\
			\label{ZEMPC_5}
			&\hat{y}(t_j|t_i) \in \mathbb{Y} , \quad j = i, i+1,\cdots, i+N-1 \\
			\label{ZEMPC_6}
			&y_z(t) \in \mathbb{Y}_z(t_j) , \quad j = i, i+1,\cdots, i+N-1
		\end{empheq}
	\end{subequations}
	where $u$, $y$ are the system input and output vector, $y_z$ is the zone tracking slack variable. $t_i$ represents the current time step, $u(t_j|t_i)$ and $\hat{y}(t_j|t_i)$ denote the system input and output at future time $t_j$ predicted at the current time $t_i$.  (\ref{ZEMPC_1}) is the objective function, where $Q$ and $R$ are diagonal weighting matrices for the zone tracking objective and the irrigation amount respectively. The control objective is to drive the system into a particular operating zone while minimizing the irrigation amount at the same time. (\ref{ZEMPC_2}) represents the two-layer NN framework discussed in the previous section, which is used to predict the system dynamics under the selected input trajectory over the control horizon $N$. (\ref{ZEMPC_3}) defines the initial state $y(t_i)$ at time $t_i$. (\ref{ZEMPC_4}) - (\ref{ZEMPC_6}) are the state, output and the zone tracking constraints, where $\mathbb{U}$, $\mathbb{Y}$, and $\mathbb{Y}_z$ are compact sets. The time-variant target zone $\mathbb{Y}_z(t_j)$ is a subset of $\mathbb{Y}$ and is defined as follow: 
	\begin{subequations}
		\label{Shrk_Z}
		\begin{empheq}{align}
			\mathbb{Y}_z(t_j) &= \{y(t)|\underline{y}(t_j) \leq y(t_j) \leq \overline{y}(t_j)\}, \quad j = i, i+1,\cdots, i+N-1 \\
			\underline{y}(t_j) &= \min{\Big[\underline{y}(t_i) \cdot e^{\mu \cdot \frac{t_j - t_i}{N}}, \underline{Y}\Big]} \\
			\overline{y}(t_j) &= \max{\Big[\overline{y}(t_i) \cdot e^{-\mu \cdot \frac{t_j - t_i}{N}}, \overline{Y}\Big]}
		\end{empheq}
	\end{subequations}
	where $\underline{y}(t_j)$ and $\overline{y}(t_j)$ represent the lower and upper bound of the target zone at time step $t_j$. The boundaries of the target zone varies over time exponentially from their initial value $\underline{y}(t_i)$ and $\overline{y}(t_i)$ until they are converged to the final value $\underline{Y}$ and $\overline{Y}$, where $\underline{Y} \leq \overline{Y}$. $\mu$ is a non-negative scalar that determines the shrinking speed of the boundaries. Larger $\mu$ leads to more aggressive shrinkage and vise versa. When $\mu = 0$, the target zone becomes time-invariant (i.e. $\underline{y}(t_j) = \underline{y}(t_i)$, $\overline{y}(t_j) =\overline{y}(t_i)$). In this work, $\underline{y}(t_i) = 0.18$ and $\overline{y}(t_i) = 0.23$ are used.
	
	Note that the magnitude of the system input $u$ is on the scale of $10^{-8}$ to $10^{-5}$, which poses numerical issues for solving optimizations problems. Thus $u$ is scaled before feeding to the optimization solver. The scaled input $u_{scaled}$ takes values between $0$ and $1$. Unless specified otherwise, $Q = 4000$, $R = 100$ are employed. 
	
	$N = 20$ is used in this work. The sample time $\Delta$ of the optimization is $2$ hours (i.e. $\Delta = 7200 $s), meaning that at a given time instant $i$, the agro-hydrological dynamics in the next $40$ hours will be optimized ($u^* = [u^*(t_i|t_i), u^*(t_i+1|t_i), \cdots, u^*(t_i+N|t_i)]$). Following the receding horizon control strategy, only the optimal input at the first step $u^*(t_i|t_i)$ will be applied to the system. The optimization problem is solved repeatedly over the horizon of interest $N_{sim} = 60$, which optimizes the system dynamics over a $5$ days duration.

	\section{ZMPC Simulation Results}	
	\label{results}
	ZMPC simulation results under various operating conditions are presented in this section. The same initial condition is employed for all simulations for fair comparisons. To be more specific,  $x_0 = -0.2\; [m]$ is the initial state at the rooting depth and is employed for the discretized Richards equation, while the corresponding system output $y_0 = 0.266\; [m^3/m^3]$ is used for the NN models. First, in Section \ref{Control_vali}, the control performance of the proposed framework is compared with the discretized Richards equation and the single LSTM discussed in previous sections. Simulations are performed under the controller defined by (\ref{ZEMPC}) with a time-invariant target zone ($\mu = 0$) in the presence of process and/or measurement noises. Then the performance of the two-layer-NN-based ZMPC with a shrinking target zone in the presence of significant noise and weather disturbance is presented in Section \ref{Shrk_Z_R}. The effects of tuning parameters $\mu$, $\underline{Y}$, and $\overline{Y}$ in ZMPC performance are investigated. The basic ZMPC with a time-invariant target zone is employed as the benchmark controller for performance comparison.
	\subsection{Optimization Performance Validation}
	\label{Control_vali}
\begin{figure}
		\centering
		\includegraphics[width=0.9\columnwidth]{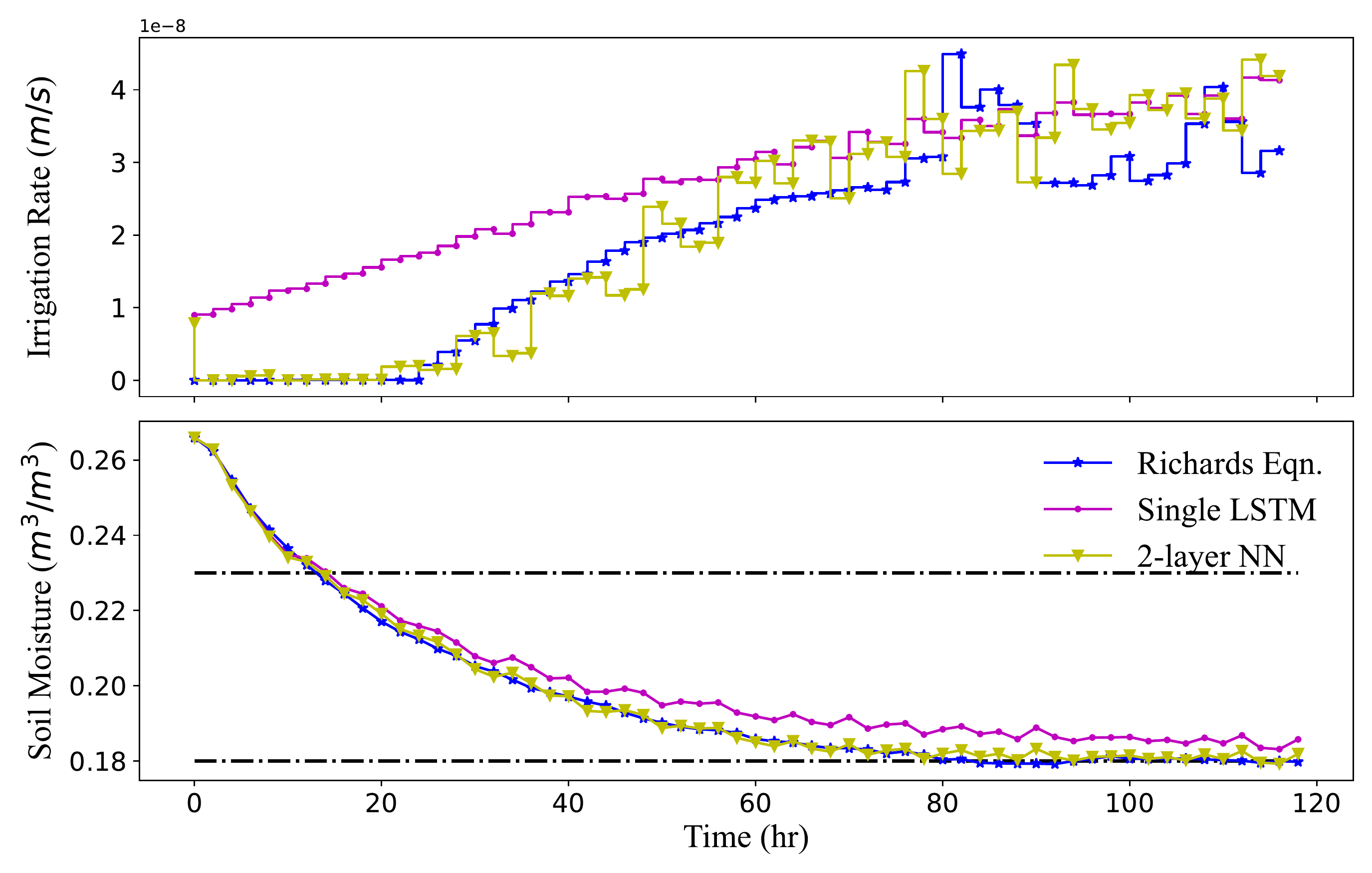}
		\caption{Optimal control strategy and the corresponding output response of ZMPC in the presence of 2\% process noise. Discretized Richards equation (star marker), the single LSTM (dot marker), the two-layer NN framework (triangle marker).}
		\label{RLP_F}
	\end{figure}

\begin{figure}
		\centering
		\includegraphics[width=0.9\columnwidth]{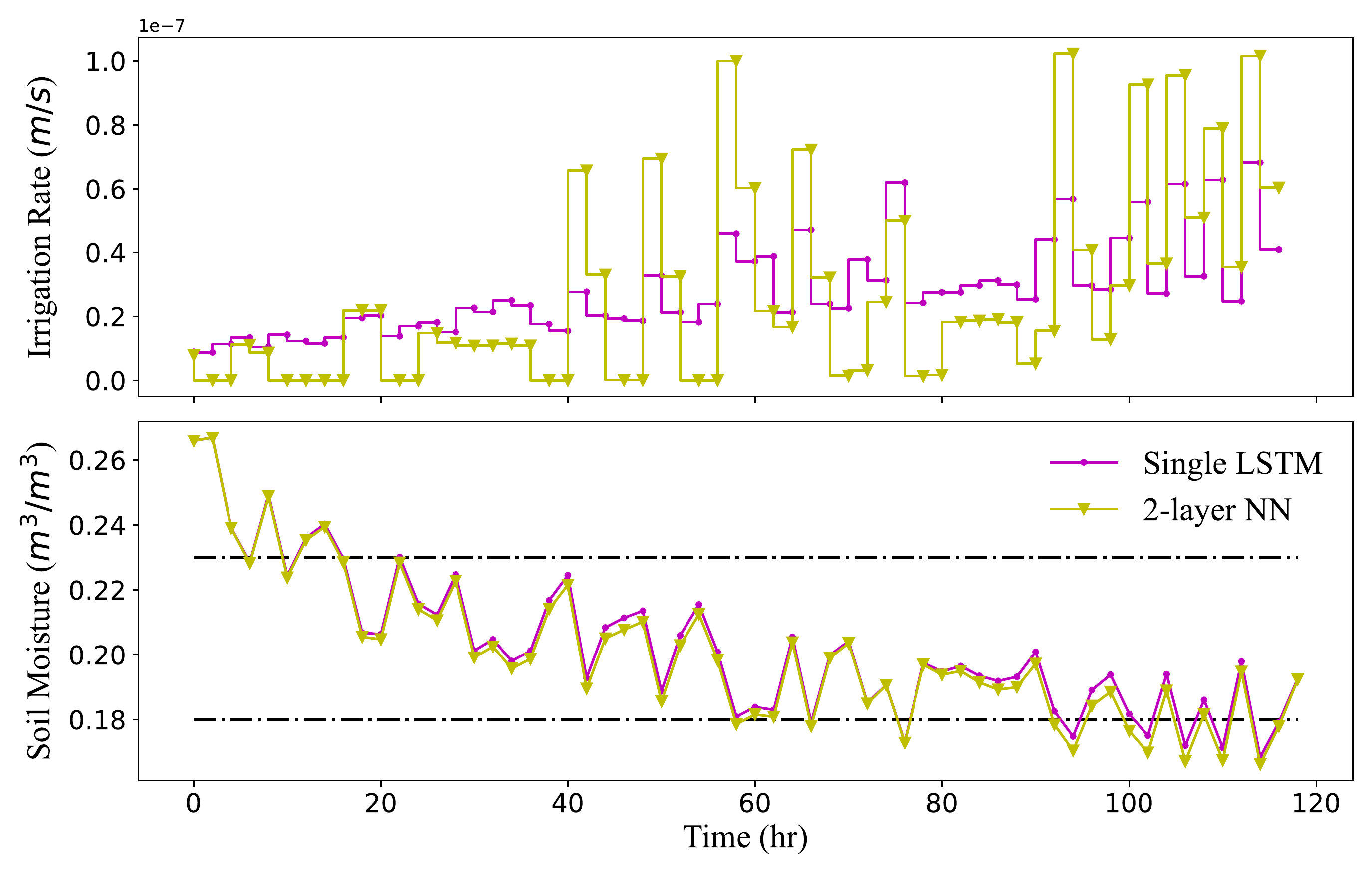}
		\caption{Optimal control strategy and the corresponding output response of ZMPC in the presence of 5\% process noise and 5\% measurement noise.}
		\label{LP_F}
	\end{figure}

	In this section the performance of the proposed framework in ZMPC applications is validated by comparing the simulation results with those obtained based on the discretized Richards equation and the single LSTM. Table \ref{RLP} summarizes the average computational time for solving the optimization problem, the economic performance of ZMPC simulations, and the percentage difference in economic performance based on different models. The economic performance is measured by the accumulated irrigation amount $I_T $ in millimeters, which is calculated as follows:
	\[I_T = \sum_{t=0}^{N_{sim}}u(t) \cdot \Delta \cdot 1000\]
	where $u(t)$ is the irrigation rate at time $t$ with the unit $m/s$. The percentage difference in the accumulated irrigation is calculated with respect to the result obtained based on the Richards equation. The optimal control strategy and the corresponding output trajectories obtained based on different models are presented in Figure \ref{RLP_F}. 
	The results shown in Table \ref{RLP} and Figure \ref{RLP_F} are obtained in the presence of 2\% process noise. Figure \ref{LP_F} presents the optimal system input and output trajectories obtained with 5\% process and 5\% measurement noise based on the single LSTM and the proposed framework. The colors and markers employed in Figure \ref{LP_F} are the same as those in Figure \ref{RLP_F}. Table \ref{LP} summarizes more ZMPC simulation results based the single LSTM and the proposed framework in the presence of process and measurement noise. 
	
	\begin{table}[!t]
		\small
		\centering
		\caption{ZMPC simulation results in the presence of 2\% process noise.}	
		\renewcommand\arraystretch{2}
		\label{RLP}
		\tabcolsep 10pt
		
		\begin{tabular*}{0.6\textwidth}{cccc}\hline
			Model & Avg. Time (s) & $I_T$ (mm) & \% Diff. in $I_T$\\ \hline
			Richards Eqn. & 493 & 8.33 & -\\
			Single LSTM & 12.5 & 11.8 & 41.4\\ 
			two-layer NN & 35.4  & 9.05 & 8.66\\ \hline
		\end{tabular*}
	\end{table}	
	
	\begin{table}[!t]
		\small
		\centering
		\caption{ZMPC simulation results in the presence of process noise (P.N.) and measurement noise (M.N.).}	
		\renewcommand\arraystretch{2}
		\label{LP}
		\tabcolsep 10pt
		
		\begin{tabular*}{0.44\textwidth}{c|cc}\hline 
			Noise & Model & $I_T$ (mm)\\ \hline\hline
			\multirow{2}{4.5em}{2\% M.N.} & Single LSTM & 11.6\\ 
			& two-layer NN  & 8.74\\ \hline
			\multirow{2}{4.5em}{5\% P.N. } & Single LSTM & 12.0\\ 
			& two-layer NN  & 9.54\\ \hline
			\multirow{2}{4.5em}{5\% M.N. } & Single LSTM & 12.0\\ 
			& two-layer NN  & 9.91\\ \hline
			\multirow{2}{5em}{2\% P.N. \& 2\% M.N.} & Single LSTM & 11.7\\ 
			& two-layer NN  & 9.46\\ \hline
			\multirow{2}{5em}{5\% P.N. \& 5\% M.N.} & Single LSTM & 12.1\\ 
			& two-layer NN  & 11.2\\ \hline
		\end{tabular*}
	\end{table}	
	
	As presented in Table \ref{RLP}, both the single LSTM and the proposed framework help to reduce the computational cost significantly. The computational cost reduction achieved by the single LSTM is more significant compare to the two-layer NN due to the simplicity of its structure. With only 8.66\% deviation, the economic performance of controller using the two-layer NN is similar to that using the Richards equation, which is significantly better compared to that obtained with the single LSTM. The trajectories in Figure \ref{RLP_F} indicate the same, as the optimal control law and the system response of the two-layer NN and the Richards equations are highly overlapped. Based on the Richards equation, no irrigation is applied for the first a few steps and starts to increase slowly over time. The two-layer NN leads to similar control strategy with the minor differences caused by the plant-model-mismatch. As for the single LSTM, the irrigation policy is less efficient, which is nonzero over the entire horizon. The output trajectories converge to the lower bound of the target zone for simulations based on the Richards equation and the proposed framework. The responses are reasonable as the economic objective is to reduce the amount of irrigation required. On the other hand, for the single-LSTM-based ZMPC, slower convergence and offsets with respect to the lower bound of the target zone is observed in the optimal output trajectory. 
	
	From Tables \ref{RLP} and \ref{LP}, it is observed that compared to process noise,  measurement noise has stronger impacts on the simulation results. As shown in Table \ref{LP}, except for the case that considered 5\% process noise and 5\% measurement noise, control with the two-layer NN has significantly better economic performances compared to that using the single LSTM. As shown in Figure \ref{LP_F}, in the presence of 5\% process noise and 5\% measurement noise, the two-layer NN provides more aggressive control policy to capture the effect of the noise compare to the single LSTM. However, the impact of noise in this case is more significant compare to the control action due to the slow-responding nature of the agro-hydrological system, as the output trajectories obtained based on the two models are similar.
	To summarize, the proposed framework is shown to have better performances compared to the single LSTM in ZMPC applications. However, some minor violations of the target zone is observed at the end of the horizon of interest (Figure \ref{LP_F}) due to the effect of noise, which is not ideal as low soil moisture may cause severe consequences in crop growing. Furthermore, precipitation is an essential disturbance in real-world applications and should be considered. In the following section, we investigate the performance of a shrinking target zone in capturing the effect of significant noise and weather disturbances.
	
	\subsection{Shrinking Zone v.s. Time-invariant Zone}
	\label{Shrk_Z_R}
	In this section, the performance of the proposed shrinking-zone ZMPC and the ZMPC with time-invariant zone (referred to as the basic ZMPC in the following text) is compared in the presence of 5\% measurement noise and weather disturbance. The controller is assumed to have information regarding the weather through weather forecasting with some uncertainties. In this work we consider 20\% error in weather forecasting. Various parameter combinations are investigated. In Section \ref{Y}, $\mu$ is kept constant and the effect of $\underline{Y}$ and $\overline{Y}$ discussed. The opposite is done in Section \ref{Mu}, where $\underline{Y}$ and $\overline{Y}$ are kept constant with varying $\mu$. 
	\subsubsection{Tuning $\underline{Y}$ and $\overline{Y}$}
	\label{Y}
	\begin{table}[!t]
		\small
		\centering
		\caption{ZMPC simulation results with different $\underline{Y}$ and $\overline{Y}$}	
		\renewcommand\arraystretch{2}
		\label{Shrk_Z_B_L_T}
		\tabcolsep 10pt
		
		\begin{tabular*}{0.95\textwidth}{ccccc}\hline
			Target Zone & $I_T$ (mm) & \% Diff. - $I_T$ & Zone Viol. ($m^3/m^3$) & \% Diff. - Zone Viol.\\ \hline
			Time Inv. Z. & 6.16 & - & 0.007 & -\\
			$\underline{Y} = \overline{Y}, \mu = 1$ & 9.91 & 61.0 & 0.0005 & 93.1\\   
			$\underline{Y} < \overline{Y}, \mu = 1$  & 9.53 & 54.9 & 0.001 & 85.1\\ \hline
		\end{tabular*}
	\end{table}		
	
	\begin{figure}
		\centering
		\includegraphics[width=0.6\columnwidth]{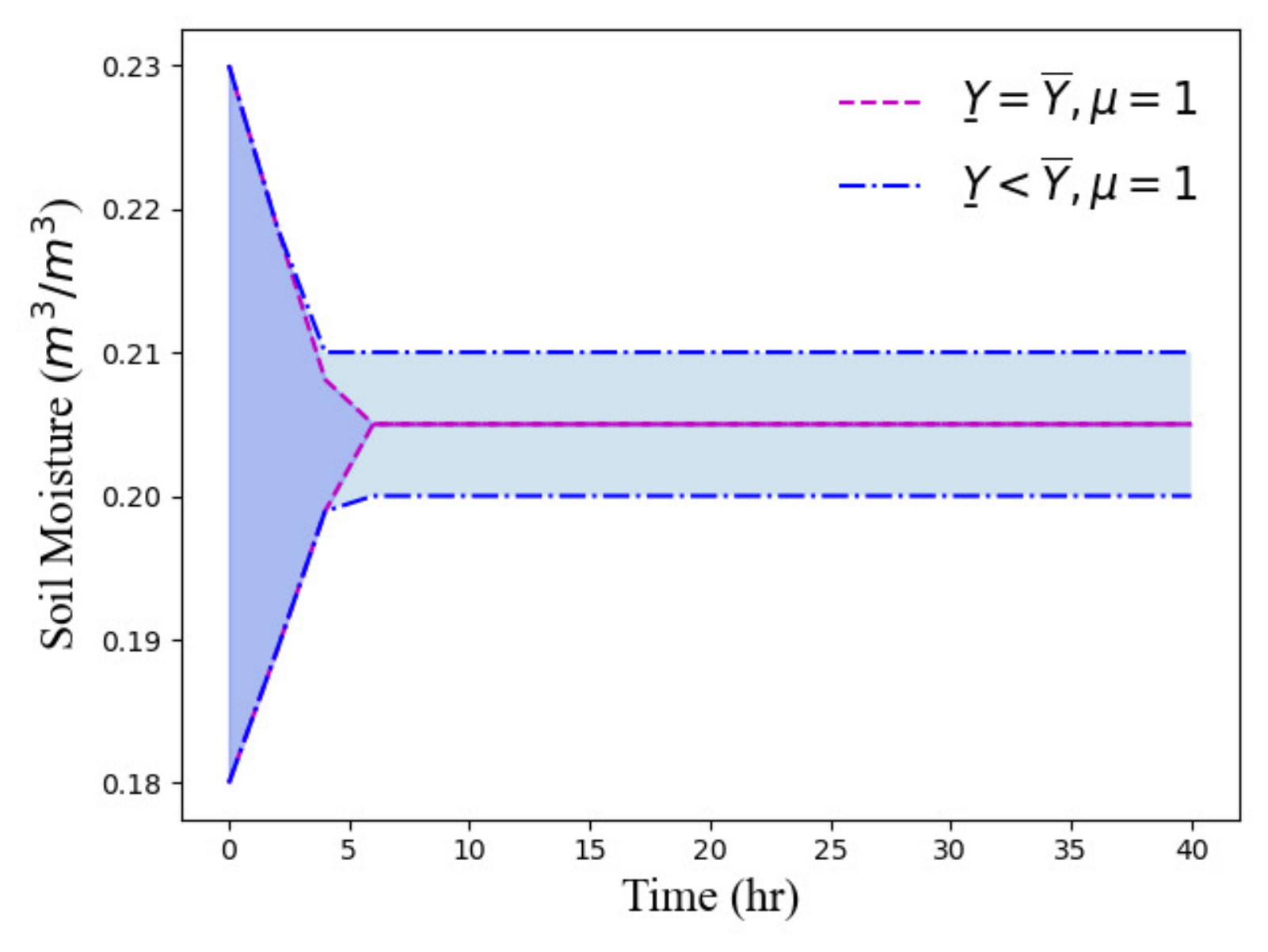}
		\caption{Shrinking zones with different $\underline{Y}$ and $\overline{Y}$.}
		\label{Shrk_Z_B_L}
	\end{figure}
	
	\begin{figure}
		\centering
		\includegraphics[width=0.95\columnwidth]{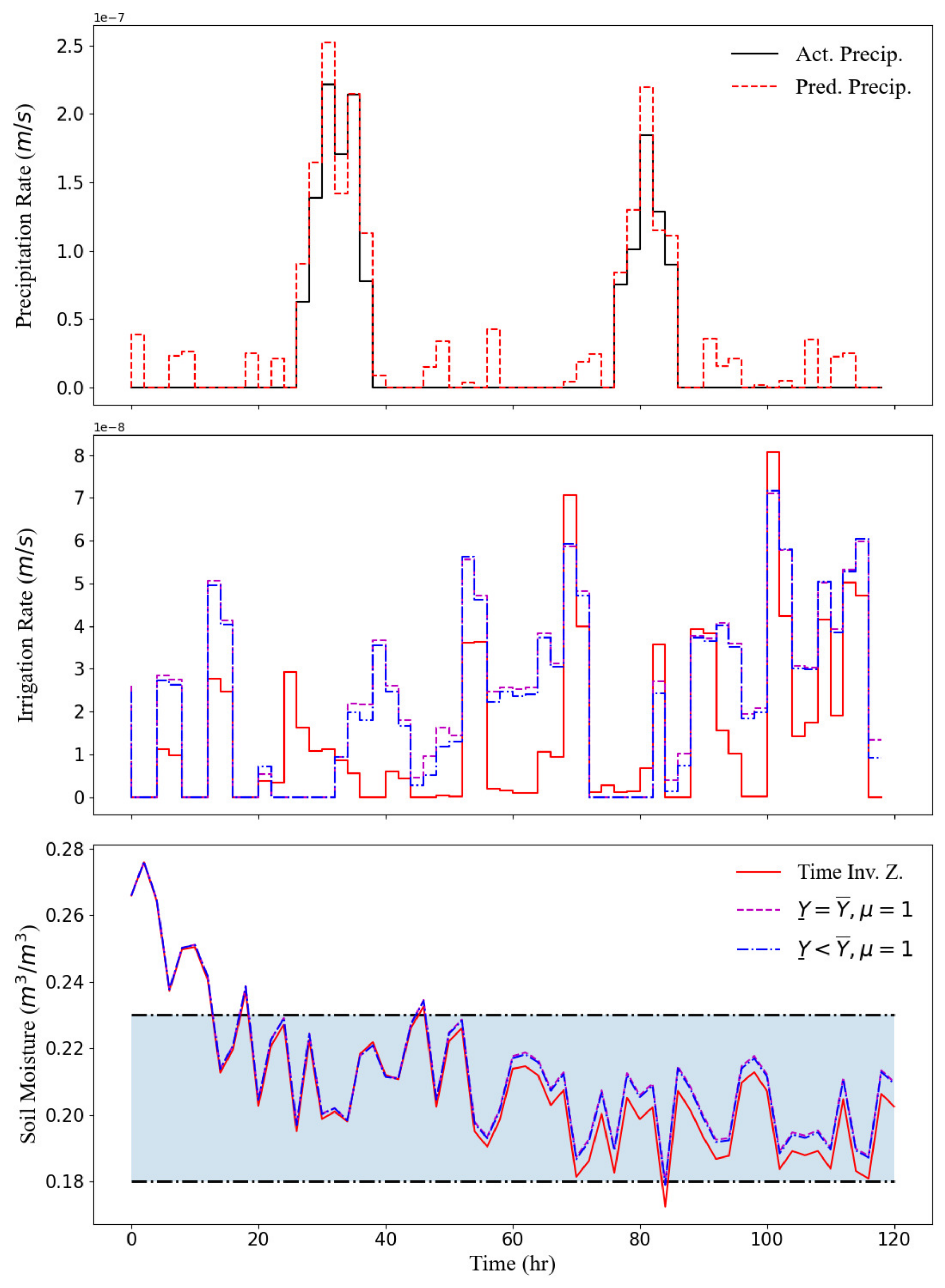}
		\caption{The precipitation data (top), optimal control strategies (middle), and corresponding output response (bottom) of ZMPC with different $\underline{Y}$ and $\overline{Y}$.}
		\label{Shrk_Z_BL_R}
	\end{figure}
	
	Figure \ref{Shrk_Z_B_L} presents two sets of shrinking zones defined with different $\underline{Y}$ and $\overline{Y}$. The darker zone bounded by dashed boundaries has $\underline{Y} = \overline{Y} = 0.205$, meaning that the target zone shrinks to the center point of the original zone. The lighter zone bounded by dashed-dotted boundaries terminates to a zone with $\underline{Y} = 0.20$, $ \overline{Y} = 0.21$. For both controllers, $\mu = 1$ is used. The ZMPC results are shown in Figure \ref{Shrk_Z_BL_R}. 
	The results obtained based on the basic ZMPC (i.e. $\underline{Y} = 0.18$, $ \overline{Y} = 0.23$) is employed as the performance benchmark. The results obtained based on the two shrinking zone design are presented with the same line style as those used in Figure \ref{Shrk_Z_B_L}. The economic and zone tracking performance of the ZMPC controllers are summarized in Table \ref{Shrk_Z_B_L_T}. The percentage difference of the total amount of irrigation and zone tracking performances are calculated based on the basic ZMPC controller.  
	
	From Figure \ref{Shrk_Z_BL_R}, it is observed that with a time-invariant zone, the controller is able to capture the effect of the weather disturbance in general. When precipitation is foreseen, the irrigation amount is reduced. However, minor violation of the target zone at the lower bound is still observed due to the presence of noise. As presented in the figure, the usage of shrinking zone helps to avoid the violation. Instead of oscillating around the lower bound of the target zone, the output trajectories converge toward the center of the zone. This provides more robustness in the controller in the presence of significant noise. It is also noticed however, the employment of shrinking zones has significantly affect the total amount of irrigation. As summarized in Table \ref{Shrk_Z_B_L_T}, the two shrinking zones considered increase the accumulated irrigation amount by 61.0\% and 54.9\% respectively. Figure \ref{Shrk_Z_BL_R} and Table \ref{Shrk_Z_B_T} indicate that the output responses based on the two shrinking zone designs are very similar, while the design that terminates to a zone performs better economically. Thus, in the following section we investigate the effect of $\mu$ with $\underline{Y} = 0.20$, $ \overline{Y} = 0.21$.
	
	\subsubsection{Tuning $\mu$}
	\label{Mu}
	\begin{table}[!t]
		\small
		\centering
		\caption{ZMPC simulation results with different $\mu$}	
		\renewcommand\arraystretch{2}
		\label{Shrk_Z_B_T}
		\tabcolsep 10pt
		
		\begin{tabular*}{0.9\textwidth}{ccccc}\hline
			Target Zone & $I_T$ (mm) & \% Diff. - $I_T$ & Zone Viol. ($m^3/m^3$) & \% Diff. - Zone Viol.\\ \hline
			Time Inv. Z. & 6.16 & - & 0.007 & -\\
			$\underline{Y} < \overline{Y},\mu = 1$ & 9.53 & 54.9 & 0.001 & 85.1\\
			$\underline{Y} < \overline{Y},\mu = 0.7$ & 8.45 & 37.2 & 0.003 & 62.0\\ 
			$\underline{Y} < \overline{Y},\mu = 0.5$  & 7.46 & 21.2 & 0.005 & 40.4\\ \hline
		\end{tabular*}
	\end{table}		
	
	\begin{figure}
		\centering
		\includegraphics[width=0.6\columnwidth]{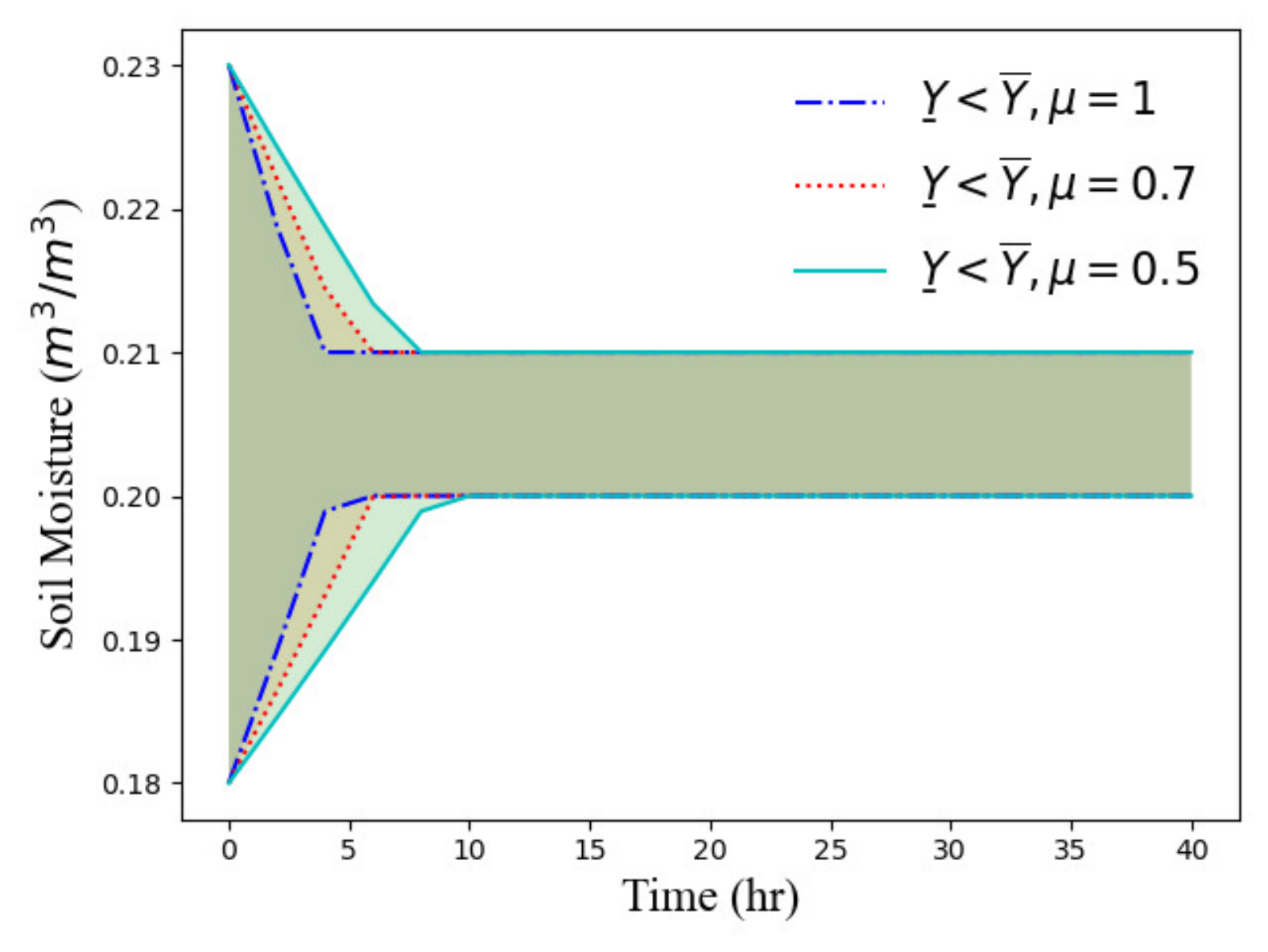}
		\caption{Shrinking zones with different $\mu$.}
		\label{Shrk_Z_B}
	\end{figure}
	
	\begin{figure}
		\centering
		\includegraphics[width=0.95\columnwidth]{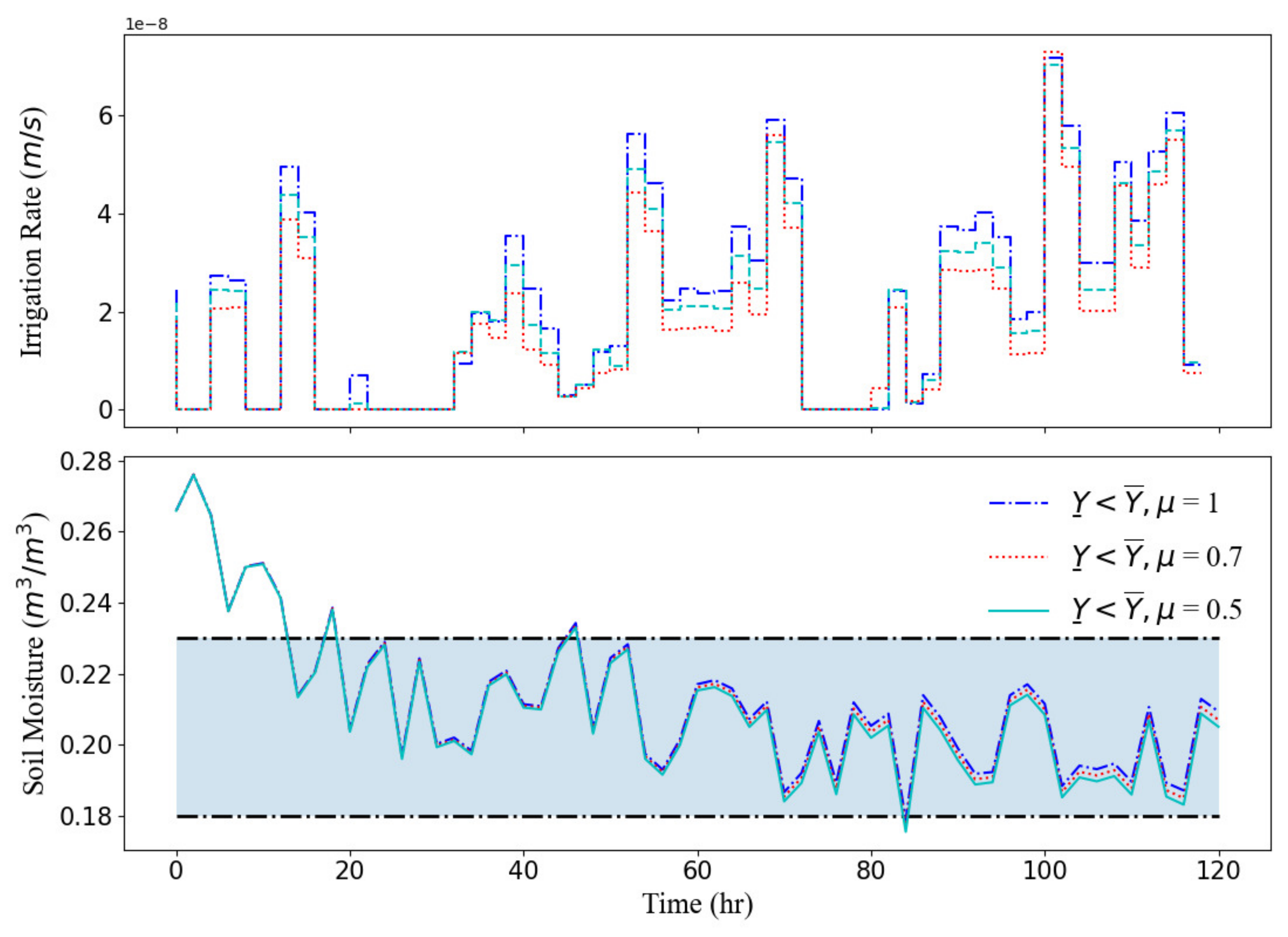}
		\caption{The optimal control strategies (top), and corresponding output response (bottom) of ZMPC with different $\mu$.}
		\label{Shrk_Z_B_R}
	\end{figure}
	
	Three shrinking zone designs with different $\mu$ are shown in Figure \ref{Shrk_Z_B}. $\mu = 1$ provides the fastest shrinking speed and thus the smallest zone. The second largest zone bounded by dotted lines has $\mu = 0.7$, while the largest zone with solid boundaries has $\mu = 0.5$. The shrinking process becomes less aggressive as $\mu$ decrease. Figure \ref{Shrk_Z_B_R} shows the optimal control strategies and the corresponding output responses of the three ZMPC controllers. The same line styles as those used in Figure \ref{Shrk_Z_B} are used in Figure \ref{Shrk_Z_B_R}. The total irrigation consumption of the three controllers is summarized in Table \ref{Shrk_Z_B_T}, where the result obtained with the basic ZMPC is again presented as the performance benchmark for consistency.
	
	From Figure \ref{Shrk_Z_B_R}, it is observed that as the magnitude of $\mu$ decrease, the output response shifts towards the lower bound of the original target zone. The target zone violation thus becomes more significant, which is consistent with the results presented in Table \ref{Shrk_Z_B_T}. It is also noticed that with reducing $\mu$ value, the total irrigation consumption reduces. In summary, in the presence of noise and disturbances, a trade off exists between the amount of irrigation applied and the zone-tracking performance. Depending on the priority of the control objectives, parameters can be tuned accordingly. If some slight zone violation is allowed, a time-invariant target zone or a shrinking zone with smaller $\mu$ and greater difference between $\underline{Y}$ and $\overline{Y}$ can be employed for better economic performance. Vice versa, a shrinking zone that shrinks more significantly over the control horizon (i.e. larger $\mu$ and narrower terminating zone) can be used if the zone-tracking objective is superior compared to the economic performance. 
	
	

	\section{Conclusion}
	\label{conclusion}
	A two-layer NN framework is proposed in this work to approximate the agro-hydrological system of interest, which is shown to be superior to a single LSTM for both open-loop prediction and closed-loop control applications. Based on the two-layer NN model, a ZMPC strategy is proposed to maintain the soil moisture content at the root zone within a certain zone. To handle the effect of noise and weather disturbance, a target zone that shrinks along the control horizon is employed in the controller. The effect of different shrinking rates and widths of the terminal zone are discussed. Through simulations, it is noticed that steeper shrinking zones help to reduce the zone violation while significantly increasing the amount of irrigation required. Mild shrinkage in the target zone leads to lower economic costs but also minor zone violations. Depending on the priority of the control objective, the hyper-parameters can be tuned accordingly.

\end{document}